\documentclass[entropy,article,accept,moreauthors]{Definitions/mdpi} 

\firstpage{1} 
\pubvolume{26}
\issuenum{12}
\articlenumber{1053}
\pubyear{2024}
\copyrightyear{2024}
\externaleditor{\textls[-25]{Academic Editors:  Andrei Khrennikov and Karl Svozil }}
\datereceived{30 October 2024 } 
\daterevised{27 November 2024} 
\dateaccepted{29 November 2024} 
\datepublished{4 December 2024} 
\hreflink{https://doi.org/10.3390/\linebreak e26121053} 
\pdfoutput=1 

\usepackage{xspace, braket, enumitem}
\usepackage{soul}

\newcommand{\tocless}[2]{\bgroup\let\addcontentsline=\nocontentsline#1{#2}\egroup}
\newcommand{\Circulant}{{\sf Circulant}\xspace}
\newcommand{\Dodis}{{\sf Dodis~et~al.}\xspace}
\newcommand{\Toeplitz}{{\sf Toeplitz}\xspace}
\newcommand{\Trevisan}{{\sf Trevisan}\xspace}
\newcommand{\VonNeumann}{{\sf Von Neumann}\xspace}
\newcommand{\Cryptomite}{{\sf{Cryptomite}}\xspace}
\newcommand{\STE}{\mathsf{STE}}
\newcommand{\CPP}{C\nolinebreak\hspace{-.05em}\raisebox{.3ex}{\bf +}\nolinebreak\hspace{-.10em}\raisebox{.3ex}{\bf +}}

\Title{Statistical Testing of Random Number Generators and Their Improvement Using Randomness Extraction}

 \TitleCitation{Statistical Testing of Random Number Generators and Their Improvement Using Randomness Extraction}

\Author{{Cameron} 
 Foreman $^{1,2,}$*\orcidA{}, Richie Yeung $^{3,4}$\orcidB{} and Florian J. Curchod $^{5}$\orcidC{}}

 \AuthorNames{Cameron Foreman, Richie Yeung and Florian J. Curchod}

\AuthorCitation{{Foreman, C.;} 
 Yeung, R.; Curchod, F.J.}

\address{%
$^{1}$ \quad Quantinuum, Partnership House, Carlisle Place, London {SW1P} 
 1BX, UK\\
$^{2}$ \quad Department of Computer Science, University College London, London {WC1E 6BT}, UK\\
$^{3}$ \quad Quantinuum, 17 Beaumont Street, Oxford OX1 2NA, UK; {richie.yeung@quantinuum.com} 
 \\
$^{4}$ \quad Department of Computer Science, University of Oxford, Wolfson Building, Parks Rd, Oxford OX1 3QG, UK \\ 
$^{5}$ \quad Quantinuum, Terrington House, 13--15 Hills Road, Cambridge CB2 1NL, UK; {florian.curchod@quantinuum.com}}

\corres{Correspondence: {cameron.foreman@quantinuum.com}}

\abstract{Random number generators (RNGs) are notoriously challenging to build and test, especially for cryptographic applications. While statistical tests cannot definitively guarantee an RNG’s output quality, they are a powerful verification tool and the only universally applicable testing method. In this work, we design, implement, and present various post-processing methods, using randomness extractors, to improve the RNG output quality and compare them through statistical testing. We begin by performing intensive tests on three RNGs---the 32-bit linear feedback shift register (LFSR), Intel's `RDSEED,’ and IDQuantique’s `Quantis'---and compare their performance. Next, we apply the different post-processing methods to each RNG and conduct further intensive testing on the processed output. To facilitate this, we introduce a comprehensive statistical testing environment, based on existing test suites, that can be parametrised for lightweight (fast) to intensive testing.}

\keyword{statistical testing; random number generation; randomness extractors; information-theoretic security} 

\begin{document}




\section{Introduction}
\label{sec:Introduction}
The notion of randomness plays an important role in numerous fields, ranging from philosophy to science. 
In science, it is used in optimisation and numerical integration (e.g.,\ using the Monte Carlo method), algorithm randomisation, or cryptography. Although~there is something universal about the concept of randomness, its definition varies substantially depending on the context in which it is used. 
In cryptography, for~example, random numbers should be \textit{{unpredictable}
}, in~the sense that they should be indistinguishable from uniformly distributed and secret ones, even by an adversary potentially possessing information about the random number generator (RNG) that the user does not have. Therefore, randomness---or unpredictability---from the perspective of the RNG user and from the perspective of a hypothetical adversary is fundamentally different. However, if~the output of the RNG exhibits patterns that are detectable by the user, then these patterns also imply predictive power from the perspective of the adversary, since the adversary needs to be considered to have at least as much information as the user. 
In this sense, unpredictability from the user's perspective is a necessary (but not sufficient) condition for the unpredictability of an adversary. 
This idea motivates the numerical testing of RNGs' outputs, which serves as a means of randomness validation, i.e.,\ to detect failures to generate~randomness.

Because numerical testing is a useful implementation check and the only universally applicable method to test different RNGs, it is an essential part of obtaining a cryptographic RNG certified by standards bodies---for~example, the~National Institute of Standards and Technology (NIST) or the Bundesamt für Sicherheit in der Informationstechnik (BSI).
This certification process ensures that the RNG has been constructed and tested following best practices. For~nearly all cryptographic applications, companies typically regard certification from a standards body as a prerequisite for RNG usage.
The NIST and BSI standards require both the detailed modelling of the underlying physical process and the numerical testing of the RNG's output statistics in order for a hardware RNG to be compliant.
This has led to the development of statistical test suites, the~best known being NIST's~\cite{rukhin2001statistical} and the Dieharder~\cite{marsaglia2008marsaglia,brown2018dieharder} suite, but~there are also others, e.g.,\ \cite{l2007testu01, ent, doty2018practrand}\@. Together, these test suites enable the comprehensive analysis of a wide array of statistical characteristics. Despite their usefulness, they are often complicated to use and their outputs are challenging to~analyse.

To address this, we design a tunable statistical testing environment {(}
$\STE$) by combining existing suites, allowing for adjustable intensity levels that balance the computational cost with effectiveness in detecting statistical bias, enabling more rigorous RNG testing than standard requirements.
We make our testing environment openly available {at} 
 \href{https://github.com/CQCL/random\_test}{https://github.com/CQCL/random\_test}\@ (accessed on 28 November 2024). 
Using the $\STE$'s most intense setting, we test the output of three different RNGs representative of those used across commercial applications, allowing us to benchmark them against each other. 
We also provide a framework to analyse the overall numerical test results generated by the $\STE$, which is not obvious~otherwise.

\begin{figure}[H]
  \includegraphics[width=0.75\linewidth]{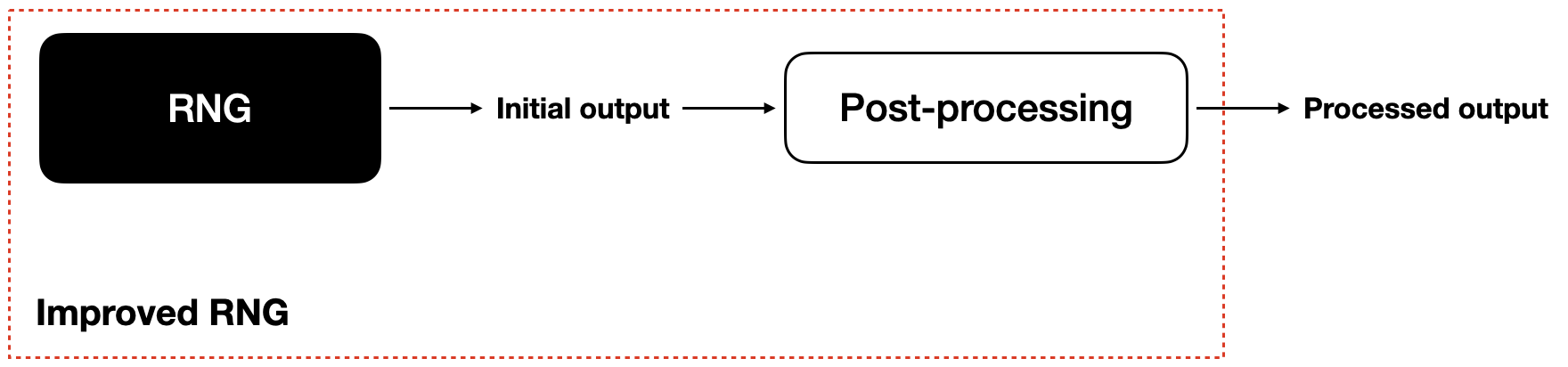}
  \caption{This figure illustrates our implementation set-up. The~black box represents one of the initial RNGs that we test, and the dashed box denotes the new---in~principle, improved---RNG with additional post-processing~applied.} 
  \label{fig:amp-setup}
\end{figure}

We then present a range of post-processing methods, utilising various randomness extractors, to~enhance the quality of each RNG's output by, for~instance, removing bias or dependencies between bits. Our set-up is illustrated in Figure~\ref{fig:amp-setup}. Each extraction method is based on distinct assumptions, which can be compared against each other. Examples of these assumptions include specific structures in the RNG’s output process, such as assuming that each output bit is generated in an identically and independently distributed (i.i.d.) manner or~the availability of a short, pre-existing (near-)perfect random bit string as a resource. To~evaluate the effectiveness of each method, we implement them using the \Cryptomite software library~\cite{cryptomite} and apply statistical testing through our $\STE$ to analyse the impact of these techniques. These assumptions must be practically justified, and~statistical testing is employed to determine whether each post-processing method succeeds in producing statistically sound~results.
\vspace{-6pt}
\unskip

\subsection{Related~Work}
\label{subsec:previous-work}
The statistical testing of RNGs has a long history, dating back to the implementation of the Diehard CD-ROM tests in the 1990s~\cite{marsaglia2008marsaglia}\@. Since then, two main research directions have emerged, which are both relevant to our work. 
First, researchers developed other test suites, such as the NIST Statistical Test Suite~\cite{rukhin2001statistical}, the~TestU01 suite~\cite{l2007testu01}, ENT~\cite{ent}, and~PractRand~\cite{doty2018practrand}, all of which we utilise for this work. 
Second, the empirical testing of specific RNGs was performed, such as~\cite{soto1999statistical, tsvetkov2011empirical}, which analysed and compared the results of statistical tests on a variety of pseudo-RNGs (PRNGs)\@. 
Other works have considered the empirical testing of so-called true-RNGs (TRNGs)---for~example,~\cite{hamburg2012analysis}, which tests the statistical properties of the entropy source in Intel's Ivy Bridge TRNG, and~\cite{jun1999intel,tsoi2007high, zhang2017640, williams2010fast, sun2018random, cho2020random}, which develop, implement, and statistically test different TRNGs. 
Recently, this has been extended to quantum RNGs (QRNGs)--- for~example,~\cite{xu2019high, o2023operation, jacak2021quantum, keshavarzian20233,hurley2020quantum}, in~which the authors extensively statistical test an ID Quantique QRNG~\cite{quantique2004quantis}\@. 
Other works give a universally applicable RNG statistical testing framework, such as~\cite{crocetti2023review, seyhan2022classification}. 

Randomness extraction also has a rich body of literature; see~\cite{shaltiel2011introduction} for an introduction. In~cryptographic randomness standards, e.g.,\ in NIST's SP 800~\cite{rukhin2001statistical}, so-called \textit{{conditioners}} are standardised, whose role is similar to that of randomness extractors ({{randomness} 
 extractors can be understood as conditioners that have information-theoretic security, i.e.,\ they do not rely on computational assumptions on the adversary}). These conditioners are the only post-processing that has been vetted for use by governing bodies and are the most commonly used as a consequence. 
To the best of our knowledge, our work represents the first attempt at comparing the effects (from a statistical perspective) of different post-processing~methods.

\subsection{Summary of~Results}
{In summary,} 
 our main results and observations are as follows.
\setlist{nolistsep}
\begin{itemize}[noitemsep]
\item We perform intense statistical testing of three different RNGs: the 32-bit Linear Feedback Shift Register (LFSR) PRNG, Intel's RDSEED TRNG, and~IDQuantique's `Quantis' QRNG.
We show the failure of two of them and provide evidence that one behaves well from a statistical perspective, extending and confirming the results of~\cite{hurley2020quantum,hamburg2012analysis}. 
\item We present and implement a variety of post-processing methods, in~the form of randomness extractors, to~improve the output quality from each of the three RNGs. The~post-processing methods form a set of four levels, where each level requires increasingly more sophisticated implementation: deterministic (level 1), seeded (level~2), two-source (level 3), and physical (level 4) extractors. Our contribution goes significantly further than the study and comparison of different types of extractors in~\cite{kwok2011comparison,ma2013postprocessing}, which focus only on deterministic or seeded extractors, respectively.
\item We present and experimentally demonstrate the implementation of a physical extractor (level 4) using a high-fidelity quantum computer. This allows us to execute a complex quantum protocol that has no classical equivalent.
\item We intensively test the statistical effect that each level of post-processing has on the output of the different RNGs. Our main observations~are as follows.
\begin{itemize}
\item RNGs that fail statistical testing without post-processing continue to fail when simple post-processing methods (level 1) are applied, although~some improvement is observed. Notably, one of these failing post-processed RNGs implements the \textit{{self-shrinking generator}} \cite{meier1994self}, which is studied for cryptographic applications.
\item All our level 2, 3, and~4 implementations successfully post-process the output of the three RNGs used from a statistical perspective, passing the statistical testing. 
\item Low-entropy sources, such as the post-processed 32-bit LFSR, can pass rigorous statistical testing when suitable post-processing is applied. While this result aligns with the existence of cryptographically secure PRNGs, the~low quality of the PRNG used highlights the limitations of statistical testing alone, i.e.,~without a precise model and justification for the unpredictability of the underlying \mbox{physical process}.
\end{itemize}
\item We make publicly available our $\STE$, a~powerful, flexible, and easy to use statistical testing environment, together with suggested settings, which provide a valuable trade-off between the intensity and computation time, and~a framework to analyse the cumulative test results.
\end{itemize}

\section{Tools and~Definitions}
\label{sec:prelimaries}
\begin{figure}[H]
  \includegraphics[width=0.75\linewidth]{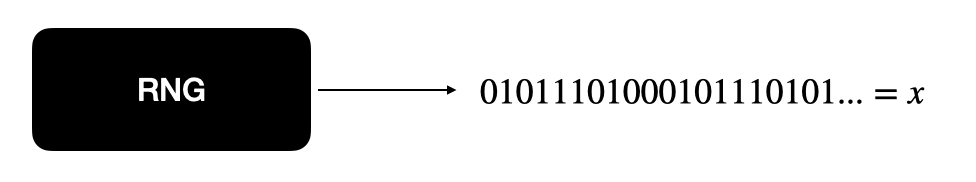}
  \caption{An illustration of the set-up that we consider. An~RNG generates a bit string $X=x$ of length $n$. 
  In this work, we first study the statistical properties of the realisation $x$ of the (random variable) $X$\@. 
  Then, we analyse the effects of different post-processing methods applied to it.}
  \label{fig:rng-setup}
\end{figure}

As discussed in the Introduction, randomness is different when considering the perspective of a user or that of an adversary. 
In this work, we study the statistical properties of different RNGs' outputs directly, without~considering any additional information that a potential adversary might obtain, such as a detailed model of the entropy source or information from invasive or non-invasive attacks.
We then examine, in~the same manner, the~effects of various post-processing methods applied to the RNG's output.
We consider the specific case of bits, i.e.,\ the RNG outputs a bit string of length $n$, denoted $X \in \{0,1\}^n$, although~one can also study RNGs whose output alphabet is larger. 
In some cases, we refer to sizes in bytes, where one byte equals eight bits. 
We denote the random variable produced by an RNG as $X$ and its specific realisation as $x$ (i.e., $X=x$). The~set-up is illustrated in Figure~\ref{fig:rng-setup}.

{The amount of randomness that a random variable $X$ has is captured by its  {min-entropy $\mathrm{H}_{\infty}(X)$.}}
\begin{Definition}[Min-entropy] \label{def:min-ent}
  The min-entropy, $k$, of~a random variable, $X \in \{ 0,1\}^n$, is defined as
\begin{align} 
      k = \mathrm{H}_{\infty}(X) = -\log_2  \max\limits_{x \in \{ 0,1\}^n} \Pr (X = x)\ . 
  \end{align}
\end{Definition}
This can be interpreted as the minimum amount of randomness, in~bits, that a~variable $X$ has when there is no side information available. 
The min-entropy rate of a random variable $X \in \{0,1\}^{n}$ is $\alpha = \mathrm{H}_{\infty}(X)/n$.
This can be interpreted as the minimum amount of randomness that $X$ has per bit, on~average. 
Since RNGs output sequentially, we generalise this definition to consider the min-entropy of the current random variable conditioned on all previously produced random variables. 
This is known as \textit{{block min-entropy}}.

\begin{Definition}[Block min-entropy] \label{def:block-min-ent}
  A set of random variables $X_i \in \{0,1\}^{n_i}$ for $i \in \mathbb{N}$ is said to have block min-entropy $k_i$, if~\begin{align} \mathrm{H}_{\infty}(X_i|X_0, X_1, ..., X_{i-1}) = -\log_2  \max\limits_{x \in \{ 0,1\}^{n_i}} \Pr (X_i = x | X_0, X_1, ..., X_{i-1}) \geq k_i ,\hspace{0.7cm} \forall i\ . \end{align}
\end{Definition}  
This can be interpreted as the minimum number of random bits that a variable $X_i$ has when conditioned on all previous random variables, indexed by $0, \ldots, i-1$.

\begin{Definition}[Statistical distance] \label{def:stat-distance}
  The statistical distance, $\Delta$, between~two random variables, $X, Z \in \{0,1\}^n$, is defined as
\begin{align}  \Delta(X,Z) = \frac{1}{2} \sum_{v \in \{ 0,1 \}^n} \lvert \Pr(X=v) - \Pr(Z=v) \rvert\ . \end{align}
\end{Definition}
This is a measure of how close to one another, or~indistinguishable from one another, two random variables are. 

\begin{Definition}[$\epsilon$-perfect randomness] \label{def:near-perfect-rand}
  A random variable $X$ on $\{0,1\}^n$ is said to be $\epsilon$-perfectly random if
\begin{align} \Delta(X, \mathrm{U}_n) \leq \epsilon\ , \end{align}
  where $\mathrm{U}_n$ is the uniform variable on $\{0,1\}^n$, i.e.,\ $\Pr(\mathrm{U}_n=u)=\frac{1}{2^n}$ for all $u \in \{0,1\}^n$. 
\end{Definition}
This definition is equivalent to saying that the variable $X$ is distinguishable from a uniform distribution with a distinguishing advantage of at most $\epsilon$, i.e.,\ a distinguisher can guess that $X$ is not uniform with a success probability of at most $\frac{1}{2}+\epsilon$\@. 
When $\epsilon = 0$, the~random variable is said to be perfectly random. 
This definition is \textit{{universally composable}}~\cite{canetti2001universally}, i.e.,\ $X$ can be used safely in other applications.

\section{Statistical~Testing}
\label{sec:stat-tests}
Statistical test suites are collections of algorithms that analyse the numerical properties of a set of random numbers to determine whether there is evidence to reject the possibility that they are uniformly distributed. 
If there is sufficient evidence to reject this possibility, a~statistical test is said to be \textit{{failed}} and the RNG output can be distinguished from the uniform distribution at some confidence level. 
The hypothesis that a random variable is uniformly distributed is known as the null hypothesis $\mathsf{H_0}$. For~an RNG producing a random variable $X \in \{0,1\}^n$, the~null hypothesis is $\mathsf{H_0}: \Delta(X,\mathrm{U}_n) = 0$. If~the null hypothesis is rejected, then the alternative hypothesis $\mathsf{H_1}: \Delta(X,\mathrm{U}_n) > 0$ is accepted.

However, a~random variable cannot be tested directly; only its realisation can---i.e.,\ the bit string $x \in \{0,1\}^n$ produced by the random variable $X$\@. 
To assess whether to accept or reject the null hypothesis, a~statistical test calculates a specific measure of $x$ (e.g., its mean), known as the \textit{{test statistic}} $t$, and~analyses how likely this test statistic is to be observed, assuming that the underlying random variable is uniform. Test statistics calculated from realisations of a uniform distribution are normally distributed, so one can calculate how likely observing certain ranges of the test statistic is by using concentration inequalities. More precisely, this likelihood is captured by a probability known as the \textit{p}-value, which is defined as follows.
\begin{Definition}[\emph{p}-value]
Given an observed test statistic $t$ obtained by calculating a measure from the realisation of a random variable $X = x \in \{0,1\}^n$ and $T$, the~(normally distributed) variable associated with all the possible measure values, the~\emph{p}-value $\mathsf{p} \in [0, 1]$ is defined as
\begin{align}
\mathsf{p} =
\Pr(T \leq t | \Delta(X, \mathrm{U}_n) = 0)\ ,
\end{align} where $\mathrm{U}_n \in \{0,1\}^n$ is uniformly distributed.
\end{Definition}

A range of \emph{p}-values is defined that provides a threshold at which the null hypothesis is rejected, i.e.,\ when the test is deemed to fail. If~a test ensures that there is, at~most, a~$1\%$ chance that it incorrectly rejects that the RNG is producing uniform random numbers ({{known} as the type 1 error---when a statistical test incorrectly rejects a true null hypothesis}), then it would, for~example, conclude failure if $\mathsf{p} \notin [0.01,1]$\@. This threshold for failure is on one tail only, so it only fails test statistics that are sufficiently biased away from the expected value in one direction. 
More generally, tests are two-tailed and conclude failure if the observed \emph{p}-values are outside of a sufficiently large interval---for~example, if~$\mathsf{p} \notin [0.005, 0.995]$. 

The failure of numerous statistical tests is a strong indicator that an RNG is not producing (near-)perfect random numbers, as~its output can be distinguished from the uniform distribution with a high probability. For~example, if~all statistical tests performed on the RNG are independent, the~probability that the null hypothesis is accepted given that the alternative hypothesis is true (known as the type 2 error) is $\mathsf{p_{type 2}}= \mathsf{p^{test1}_{type 2}} \cdot \mathsf{p^{test2}_{type 2}} \cdot \ldots \cdot \mathsf{p^{test n}_{type 2}}$, where $n$ is the number of tests~performed.

We now describe several existing statistical test suites used in this work. It is important to note that, while each suite contains multiple tests, many test outcomes are correlated. For~instance, a~source with an unusually high number of ones may fail both the monobit and poker tests. Similarly, different test suites often include the same tests with slight parameter variations, meaning that failure in one suite is likely to result in failure in another. For~example, both NIST and Dieharder include a runs~test.

\subsection{Existing Test~Suites} 
\label{subsec:existing-tests}

\subsubsection{{NIST} 
 Statistical Test Suite}
The NIST statistical test suite (SP 800-22) \cite{rukhin2001statistical} is the best known. It includes 15 tests, some with multiple sub-tests, and~passing certain tests is required for RNG certification by organisations such as NIST and BSI. During~testing, a~randomness file is split into sub-strings, with~each sub-string tested individually. Users can specify the number of sub-strings and the total bit string size to analyse, although~the guide recommends using 100 sub-strings of $10^6$ bits, requiring at least $10^8$ bits, or~12.5 MB, for~testing.
{For each test, an analysis is performed on each sub-string, and~the suite provides two results: (1) the \emph{p}-value for a statistical test on the uniformity of the distribution of results across sub-strings and~(2)~the number of sub-strings that pass each test. Both results are assessed at the 1\% significance level. We note that the individual \emph{p}-values for each sub-string in each test are not accessible to the user.} 

\subsubsection{Diehard(er) Statistical Test Suite} 
The Dieharder statistical test suite includes the 18 original Diehard tests along with additional tests, including some from the NIST suite. Like the NIST suite, it is widely used by RNG certification bodies. A~failure is determined when $\mathsf{p} \notin [0.0005, 1 - 0.0005]$, and~a test is considered 'weak' if $\mathsf{p} \in [0.0005, 0.005] \cup [1 - 0.0005, 1 - 0.005]$. This higher tolerance for poor test statistics means {that} a flawed RNG may occasionally pass Dieharder, but~failure is a strong indicator of non-uniformity. The~Dieharder tests require a large quantity of random numbers to prevent the re-use of input data, which can lead to inaccurate results. We recommend using at least 1 gigabyte (GB) of random numbers for testing. For~smaller file sizes, the test parameters can be adjusted to avoid these issues. In~our testing environment, we use the default parameters for each~test. 

\subsubsection{TestU01 Statistical Test Suite}
TestU01 is a C-based software library for conducting RNG statistical testing with pre-compiled test batteries. These batteries vary significantly in the number of tests and the amount of randomness required. For~details on the specific tests included in each battery, see~\cite{l2007testu01}. Test \emph{p}-values are reported if $\mathsf{p} \notin [0.001, 0.999]$, which we use as our failure criterion. In~our testing, we use the Alphabit, Rabbit, and~SmallCrush batteries from TestU01. To~run these tests, the input files must contain at least $2^{25}$ random bytes (approximately 35 MB). We omit the Crush and BigCrush batteries due to their long runtimes and large randomness requirements, although~they can be run within our statistical testing~environment. 

\subsubsection{ENT Statistical Test Suite} 
The ENT test suite is a small but efficient set of six statistical tests. It has been used to demonstrate bias in a commercial quantum RNG by consistently failing the $\chi^2$ test~\cite{hurley2017certifiably} (we replicate these results with an independently acquired RNG; {see} 
 Table 3. ENT outputs test statistics without providing a pass/fail threshold, so we assess failure based on the criteria in Table~3 of~\cite{ortiz2018heartbeats}. Although~there is no specific guidance on the required input sizes, we found that the tests produced unreliable results with inputs smaller than $0.5$ GB.

\subsubsection{PractRand Statistical Test Suite} 
PractRand is a \CPP\: library of statistical tests designed for practicality---they are efficient, user-friendly, and~capable of detecting significant biases in RNGs. According to its documentation, it runs faster than most test suites (which we confirm; {see} Table 2), offers unique interfacing, has no theoretical maximum input length, and~includes some original tests.
PractRand tests input files based on size, examining subsets of $2^{24+x}$ bytes for $x \in \mathbb{N}$, with~more tests performed as $x$ increases. In~our testing, we limit the maximum test size to $2^{32}$ bytes (approximately 4.3 GB). For~full details and comparisons with other test suites, see~\cite{doty2018practrand}.
PractRand uses various \emph{p}-value ranges, including ``unusual'', ``mildly suspicious'', ``suspicious'', ``suspect'', and ``fail''. A~failure occurs when $\mathsf{p} \notin [10^{-11}, 1 - 10^{-11}]$. 

\subsection{Our Statistical Testing~Environment}
The interfacing code for our $\STE$ can be downloaded {at} \url{https://github.com/CQCL/random\_test} (accessed on 28 November 2024). 
We offer three testing modes, \textit{{Light}}, \textit{{Recommended}}, and~\textit{{All}}, which can be executed using the commands \texttt{run\_light}, \texttt{run\_recommended}, and~\texttt{run\_all}, respectively. The~NIST statistical test suite is not included in these commands due to its need for user prompts, but~it can be run separately within the environment. We believe that the \textit{{Recommended}} setting strikes a good balance between the computational (and environmental) cost and rigor, exceeding the standard testing required by certification bodies. All results in this work can be replicated using the $\STE$ or by downloading, configuring, and~running the relevant statistical test suites~independently.

\subsubsection{Suggested~Settings}
\label{subsec:suggested-tests}
We now propose the recommended settings for statistical testing using our $\STE$, based on insights gained during this research. The~runtimes are averaged over 10 executions using a 10 Gbit file, except~for the NIST suite, where a 100 Mbit file is tested, in~accordance with the user guidelines. All testing was conducted on a Dell Precision 7540 laptop with 16 GB of RAM and a~2.3 GHz Intel I9-9880H processor,~running the Ubuntu 20.04 operating~system. {All the test runtimes are given in} Tables~\ref{table1} and \ref{table:test-speeds}. 

\begin{table}[H]
\caption{{This} 
 table details our settings for light, recommended, and all statistical testing using the code provided. A~`Y' in a specific column indicates that the associated test suite of this column is included in the~setting.}
\begin{adjustwidth}{-\extralength}{0cm}\label{table1}
		\newcolumntype{C}{>{\centering\arraybackslash}X}
  \begin{tabularx}{\fulllength}{cCCCCCCCcC}
    \toprule
    \begin{tabular}[c]{@{}c@{}}\textbf{Test}\\ \textbf{Mode}\end{tabular}   & \begin{tabular}[c]{@{}c@{}}\textbf{NIST}\\ \textbf{(15)}\end{tabular} & \begin{tabular}[c]{@{}c@{}}\textbf{Diehard}\\ \textbf{(18)}\end{tabular} & \begin{tabular}[c]{@{}c@{}}\textbf{ENT}\\ \textbf{(6)}\end{tabular} & \begin{tabular}[c]{@{}c@{}}\textbf{SmallCrush}\\ \textbf{(15)}\end{tabular} & \begin{tabular}[c]{@{}c@{}}\textbf{Alphabit}\\ \textbf{(17)}\end{tabular} & \begin{tabular}[c]{@{}c@{}}\textbf{Rabbit}\\ \textbf{(40)}\end{tabular} & \begin{tabular}[c]{@{}c@{}}\textbf{PractRand}\\ \textbf{(920)}\end{tabular} & \textbf{\begin{tabular}[c]{@{}c@{}}\textbf{Total} \\ \textbf{Runtime}\end{tabular}} & \textbf{\begin{tabular}[c]{@{}c@{}}\textbf{Total} \\ \textbf{Tests}\end{tabular}} \\ \midrule
    Light       &                                                     &                                                        & Y                                                 & Y                                                         &                                                         &                                                       & Y                                                         & {4 m 44 s} 
                                                      & 941                                                    \\ \midrule
    Recommended & Y                                                  & Y                                                     & Y                                                 &                                                           &                                                         & Y                                                     & Y                                                         & 114 m 31 s                                                      & 999                                                    \\ \midrule
    All & Y                                                  & Y                                                     & Y                                                 & Y                                                          & Y                                                        & Y                                                     & Y                                                         & 127 m 41 s                                                      & 1015                                                    \\ \bottomrule
  \end{tabularx}
  \end{adjustwidth}
  \label{table:test-recommendations}
\end{table}
\unskip

\begin{table}[H]
 \caption{{This} 
 table gives the average runtimes of all statistical test suites contained in our statistical testing environment. 
  The average is taken when running each test suite 10 times on independent inputs. 
  For the NIST test suite, the~runtime relates to testing a 100 Mbit file.
  For all other test suites, the~runtime is for a 10 Gbit file.} \label{table:test-speeds}
  \newcolumntype{C}{>{\centering\arraybackslash}X}
\begin{tabularx}{\textwidth}{Cccccccc}
  \toprule
                  & \begin{tabular}[c]{@{}c@{}}\textbf{NIST}\\ \textbf{(15)}\end{tabular} & \begin{tabular}[c]{@{}c@{}}\textbf{Diehard}\\ \textbf{(18)}\end{tabular} & \begin{tabular}[c]{@{}c@{}}\textbf{ENT}\\ \textbf{(6)}\end{tabular} & \begin{tabular}[c]{@{}c@{}}\textbf{SmallCrush}\\ \textbf{(15)}\end{tabular} & \begin{tabular}[c]{@{}c@{}}\textbf{Alphabit}\\ \textbf{(17)}\end{tabular} & \begin{tabular}[c]{@{}c@{}}\textbf{Rabbit}\\ \textbf{(40)}\end{tabular} & \begin{tabular}[c]{@{}c@{}}\textbf{PractRand}\\ \textbf{(920)}\end{tabular} \\ \midrule
  Average Runtime & 37 m 3 s & 18 m 12 s & 1 m 24 s & 0 m 32 s & 12 m 38 s & 55 m 4 s & 2 m 48 s \\ \bottomrule
  \end{tabularx}
\end{table}

\paragraph{{Light}
}
Our suggested light test mode, executed with the command \texttt{run\_light}, includes ENT, SmallCrush, and~PractRand, running in under 5 min and covering approximately \mbox{941 tests.} Our numerical analysis shows that this set of tests is sufficient to detect failures in generating uniform randomness for RNGs that fail (see Sections \ref{sec:rng-analysis} and \ref{sec:rand-amp-processes}).

\paragraph{Recommended} 
Our recommended setting, executed with the command \texttt{run\_recommended}, includes most of the light mode test suites, replacing TestU01 SmallCrush with Rabbit and~adding the NIST and Diehard tests. These additional tests increase the runtime to approximately 2 h and bring the total number of tests to 999. The~full statistical test mode (\texttt{run\_all}) includes all suites, except~for SmallCrush and Alphabit, which are omitted due to their significant correlations with the Rabbit tests, where only the parameters differ slightly. The~recommended suite covers all tests required by RNG certification bodies (e.g., NIST and Dieharder) while offering a more comprehensive analysis than when running these tests alone. The later sections demonstrate the value of extending beyond individual test suites, as~an RNG that passes the NIST and Dieharder tests can still show significant statistical bias when analysed with our combined STE (see \Cref{sec:rng-analysis}). 

\subsection{Shortcomings of Statistical~Testing} 
\label{subsec:shortcomings-stats-tests}

Fundamentally, statistical tests have a limited ability to validate that good random numbers are being produced by an RNG. 
They should rather be understood as a useful tool to detect a failure to generate uniform random numbers, since passing statistical tests gives no guarantee of (near-)perfect randomness. 
This is especially important in the case of cryptographic RNGs. 
For example, in~\cite{shrimpton2015provable}, a~thorough analysis of Intel's RDSEED hardware RNG is performed, and one of their conclusions is that ``RDSEED delivers truly random bits but with a security margin that becomes worrisome if an adversary can see a large number of outputs from either interface. If~he controls an unprivileged process on the same physical machine, this could happen very quickly'' ({{in this case}, the~adversary also requires control of an ``unprivileged'' process, which is a form of side information that may be difficult to obtain in practice}) (on page 4)\@. 
As we shall see next, our statistical testing results do not detect that RDSEED's output can be distinguished from~uniform.

At the implementation level, the~available software for numerous statistical test suites has been shown to have issues. For~the NIST test suite, the~list of implementation issues is extensive, so we summarise a few problems that the reader may find interesting.
Research has found significant dependencies between the tests~\cite{burciu2019systematic} and implementation issues with certain tests; for example,~\cite{hamano2007correction} found that the settings of both the Discrete Fourier Transform test and Lempel-Ziv test were wrong, and~\cite{kowalska2022revision} found an error within the probability calculations for the Overlapping Template Matching test. 
Moreover, problems with how the results are analysed have been discovered; for~example,~\cite{marton2015interpretation} found that although the NIST documentation provides guidance that the analysed RNG is random if all tests are passed, truly uniform data have a high probability (80\%) of failing at least one NIST statistical (sub-)test.
Some work has even suggested that the tests are ``harmful'' \cite{saarinennist}, namely that ``the weakest pseudo-random
number generators will easily pass these tests, promoting
false confidence in insecure systems''. 
During this work, we found an additional issue with the NIST test suite that we could not find reported elsewhere: the results showed that all tests failed whenever the CPU was being used for other computations simultaneously. 
The NIST Random Bit Generation Team have been made aware of this. 
Other test suites have also had their own reported problems, including Dieharder. In~\cite{sys2022bad}, it was found that over 50\% of the Dieharder tests generated biased null hypothesis distributions (which were expected to be uniform). 

\section{Statistical Testing of Different~RNGs}
\label{sec:rng-analysis}
In this section, we use our $\STE$ to analyse the statistical properties of the random numbers produced by some commonly used RNGs. At~this stage of our analysis, we do not apply any post-processing to the RNG's output; however, some of the RNGs that we consider already have post-processing included, in~the form of so-called conditioners or deterministic randomness extractors. Therefore, in~cases where the post-processing already exists, our statistical analysis applies to the joint system, which comprises both the randomness (or entropy) source and the existing post-processing in the device. Additionally, we use and discuss a NIST min-entropy estimation tool, which provides a min-entropy estimate for our subsequent analysis incorporating various levels of~post-processing.

The RNGs that we analyse are as follows:
\setlist{nolistsep}
\begin{itemize}[noitemsep]
  \item 32-bit LFSR: a software pseudo-RNG; 
  \item Intel RDSEED~\cite{jun1999intel}: a hardware RNG based on thermal noise, i.e.,\ a chaotic process;
  \item IDQuantique (IDQ) Quantis~\cite{quantique2004quantis}: a hardware RNG based on the quantum effect of detecting photons at the output of a semi-transparent mirror. 
\end{itemize}

Further details and descriptions of the RNGs can be found in \Cref{app:quantis-qrng}. The~LFSR remains widely used in numerical simulations, despite its well-known flaws~\cite{stkepien2013statistical}. In~this work, we primarily use it as a benchmark, serving as an example of a poor choice due to the patterns in its output and its short period. However, we will show that post-processing its output has a statistically significant impact. We note that IDQ's Quantis is marketed as an RNG for cryptographic use, with~certifications for compliance with various security and cryptographic standards. ({{IDQ} Quantis has reportedly passed certifications or government validations, including ``NIST SP800-22 Test Suite Compliance, METAS Certification, CTL Certification, multiple iTech Labs certificates, and compliance with BSI's AIS31 standard (dedicated version of Quantis)'' \cite{quantique2004quantis}}).
These RNGs are used in numerous applications and are a sample of the different types of RNGs available today. 
The statistical analysis is performed using the \texttt{run\_all} function in our statistical test environment on $10 \times 10$ Gbit files from each RNG and, similarly, using the NIST test suite performed on $10 \times 100$ Mbit files split into 100 sub-strings, each of $1$ Mbit. The~NIST min-entropy estimators~\cite{mckay2016users} are used in the non-IID setting on $10\times 1$ Mbit files. 
This analysis far exceeds that required by certification bodies, so it may be a result of independent interest. 
All testing is performed using the default parameters, unless~otherwise~stated. 

An RNG producing near-perfect randomness should pass almost all statistical tests. More concretely, we mean that the ideal RNG would fail less than 7.5 of the 4600 individual statistical tests on average. ({{This} number is the expected amount of type 2 error, i.e.,\ the expected maximum number of failed tests, given that the underlying distribution is indistinguishable from uniform. Note that we are implicitly assuming that each statistical test is independent}). The~results obtained for the three RNGs are summarised in the following Table~\ref{table:all-tests-no-postprocessing} and displayed visually {in} 
 Figure 4 (level 0).

\begin{table}[H]
\caption{{This} 
 table gives the average sum of statistical tests failed for $5 \times 10$ Gbit samples from each RNG (after testing 10 samples)\@. 
  The results are presented in this way to allow for direct comparison to later results, where only $5 \times 10$ Gbit samples are tested.   
  Due to the 32-bit LFSR failing PractRand quickly, only 635 tests were conducted (instead of the full 4600), so we rescale these results. 
  In cells with multiple entries, failed tests are on the left and \textit{suspicious} tests (when applicable) are on the right in parentheses.
  The full results can be found in \Cref{app:rng-benchmarking}\@.}
  \begin{adjustwidth}{-\extralength}{0cm}
		\newcolumntype{C}{>{\centering\arraybackslash}X}
		\begin{tabularx}{\fulllength}{CCCCCCCC}
  \toprule
  \multirow{2}{*}[0.25em]{\textbf{RNG}}      & \begin{tabular}[c]{@{}c@{}}\textbf{NIST}\\ \textbf{(75)}\end{tabular} & \begin{tabular}[c]{@{}c@{}}\textbf{Diehard}\\ \textbf{(90)}\end{tabular} & \begin{tabular}[c]{@{}c@{}}\textbf{ENT}\\ \textbf{(30)}\end{tabular} & \begin{tabular}[c]{@{}c@{}}\textbf{SmallCrush}\\ \textbf{(75)}\end{tabular} & \begin{tabular}[c]{@{}c@{}}\textbf{Alphabit}\\ \textbf{(85)}\end{tabular} & \begin{tabular}[c]{@{}c@{}}\textbf{Rabbit}\\ \textbf{(200)}\end{tabular} & \begin{tabular}[c]{@{}c@{}}\textbf{PractRand}\\ \textbf{(4600)}\end{tabular} \\ \midrule
  32-bit LFSR & 10                                                    & 40  (3)                                                  & 5                                                  & 51                                                        & 73                                                       & 131                                                    & 855  (167)                                              \\ \midrule
  RDSEED      & 0                                                    & 0  (4)                                                   & 0                                                  & 1                                                         & 0                                                        & 1                                                      & 0  (7)                                                   \\ \midrule
  IDQ Quantis & 0                                                    & 0  (3)                                                   & 5                                                  & 0                                                         & 17                                                       & 25                                                     & 3  (15)                                                    \\ \bottomrule
  \end{tabularx}
  \end{adjustwidth}
  \label{table:all-tests-no-postprocessing}
\end{table}

In the statistical tests, the~RDSEED RNG performs the best, failing the fewest tests, and the~32-bit LFSR performs the worst, failing the most. The~poor performance of the LFSR is likely due to its periodicity, since bits repeat every $2^{32} - 1$ ($4.3$ Gbits), and this is less than the size of the files tested. The~IDQ Quantis device performs well in the NIST and Diehard tests but fails an ENT test and several tests in TestU01's Rabbit and Alphabit suites. These observations reproduce (and add confidence to) the results of previous work~\cite{hurley2020quantum}.
These results, especially for IDQ's device, exhibit the need to go beyond the requirements of certification bodies for statistical testing, with~additional tests providing a noticeable advantage in detecting~failures.

The NIST min-entropy estimators~\cite{mckay2016users} are a collection of algorithms that give a standardised means of estimating the min-entropy (as defined in \Cref{def:min-ent}) of an RNG's output. These estimators are useful both in evaluating the entropy generation of the RNG and in calculating a min-entropy bound, which we later use to determine the parameters for randomness extractors. Although~the estimators are designed to test entropy sources without post-processing, this is not feasible in the case of RDSEED or the IDQ Quantis~device.

In Table~\ref{table:min-ent-est}, the~NIST min-entropy estimator per byte is the average observed per-byte min-entropy calculated by the NIST min-entropy estimator tool, while $\overline{\mathsf{est}}$ is the per-bit average. 
The sample standard deviation, $\sigma$, reflects the variability in the different test results calculated using \Cref{eq:sigma}, and~$\alpha$ is a lower bound (with probability of at least $1-2^{-32}$) on the per-bit min entropy for any test sample. 
Details of this derivation can be found in \Cref{app:min-ent-bound}.

\begin{table}[H]
\caption{{This} 
 table shows the average NIST min-entropy estimator, the~sample standard deviation, and a lower bound for min-entropy/bit for each RNG. 
  These results are the average of 10 tests on different 1'000'000 bit samples, each generated with significant time gaps between the generation of each test sample.
  Full results can be found in \Cref{app:min-ent-est}\@.}
  \newcolumntype{C}{>{\centering\arraybackslash}X}
\begin{tabularx}{\textwidth}{CCCCC}
  \toprule
  \multirow{4}{*}[0.65em]{\textbf{RNG} }        & \begin{tabular}[c]{@{}c@{}}\textbf{NIST}\\ \textbf{Min-Entropy} \\ \textbf{Estimator}\\ \textbf{(/byte)}\end{tabular} & \begin{tabular}[c]{@{}c@{}}\boldmath{$\overline{\mathsf{est}}$}\textbf{: NIST} \\ \textbf{Min-Entropy}\\ \textbf{Estimator}\\ \textbf{(/bit)}\end{tabular} & \begin{tabular}[c]{@{}c@{}}\boldmath{$\sigma$}\textbf{: Sample} \\ \textbf{Standard}\\ \textbf{Deviation}\\ \textbf{(/bit)}\end{tabular} & \begin{tabular}[c]{@{}c@{}} \boldmath{$\alpha$}\textbf{: Lower} \\ \textbf{Bound} \\ \textbf{Min-Entropy}\\ \textbf{(/bit)}\end{tabular} \\ \midrule
  32-bit LFSR & 6.870                                                                             & 0.859                                                                            & 0.058                                                                           & 0.453                                                                          \\ \midrule
  RDSEED      & 6.189                                                                             & 0.852                                                                            & 0.022                                                                           & 0.698                                                                          \\ \midrule
  IDQ Quantis & 7.157                                                                             & 0.895                                                                            & 0.006                                                                           & 0.853                                                                          \\ \bottomrule
  \end{tabularx}
  
  \label{table:min-ent-est}
\end{table}

The IDQ Quantis device had the highest estimated min-entropy per bit, with~a value of 0.895, although~all three RNGs had similar values. The~32-bit LFSR had the largest sample standard deviation, indicating the greatest fluctuation in the min-entropy estimates across different test samples. We note that the NIST SP800-90 series recommends min-entropy per bit of at least $1 - 2^{-32}$ for an RNG to be considered to have full entropy~\cite{buller2022discussion}. This value is significantly higher than any values that we observed. However, some of the NIST min-entropy estimator tests are known to produce significant underestimates~\cite{zhu2017analysis}, which potentially explains the large disparity between our estimates and the NIST requirement. Moreover, in~our case, underestimates are not problematic, since we desire a lower bound on the min-entropy of the RNG's~output.

\section{A Variety of Post-Processing~Methods}
\label{sec:rand-amp-processes}

Randomness extractors are mathematical algorithms that distil \textit{weakly} random bit strings ({{more} precisely, a~necessary (but not sufficient) condition for randomness extraction to be successful is that the source has some min-entropy; see \Cref{def:min-ent}\@}), in~the sense that they are not uniformly distributed, into~a near-perfect random bit string.
In this section, we present, implement, and test a variety of randomness extraction processes.
The main question that we seek to answer is whether these methods have an observed impact on the statistical properties of the RNG's output. 
The procedure that we follow is the following.
\setlist{nolistsep}
\begin{itemize}[noitemsep]
  \item[1.] We collect the output of each RNG that was tested in the previous section. 
  We call this the \textit{{initial}} output.
  \item[2.] We apply different post-processing methods, or~randomness extractors, to~this initial output to produce a new, \textit{processed} output. Each time, we precisely define and explain the underlying assumptions of the used extractors required for the extraction method to be successful. 
  These different sets of assumptions, for~each extraction method, can be compared with each other and form the different post-processing levels.
  \item[3.] We analyse the new, processed output with our $\STE$ to determine whether each extraction method had an impact from a statistical perspective. We also compare the results obtained using the different post-processing methods for~each RNG\@.
\end{itemize}

A schematic of the set-up can be found in Figure~\ref{fig:amp-setup}\@.

\subsection{Randomness Extraction~Methods}
\label{subsec:hierarchy}
We now describe the different post-processing levels that we consider in this work, i.e.,\ the types of randomness extractors that we will use to improve the different RNGs.
We consider four classes of randomness extractor, which form the different levels, each with increasingly elaborate implementations.
\setlist{nolistsep}
\begin{itemize}[noitemsep]
  \item \textbf{{Level 1: Deterministic extractors}}---This class of extractors requires certain properties of the initial output's distribution to hold, beyond~just a min-entropy assumption. An~example is the seminal \VonNeumann extractor~\cite{von1963various}, which works if every bit of the initial output is identically and independently generated (although a sufficient condition is that the input forms an exchangeable sequence)\@. 
  In practice, assumptions of this type are difficult to justify and to control. 
  \item \textbf{{Level 2: Seeded extractors}}---These extractors require a second string, called a \textit{seed}, of~independent and (near-)perfectly random bits as the resource. This seed needs to be carefully generated and~can lead to problems if, for~example, it is not generated independently of the initial output of the RNG ({{this} could happen, for~example, if~the seed is generated whilst sharing the same environment as the RNG or by an adversary}) or if it has poor statistical properties. At~a fundamental level, seeded extractors are unsatisfying as there is circularity in having to generate near-perfect randomness as a resource to build an RNG.
  \item \textbf{{Level 3: Two-source extractors}}---These extractors are a generalisation of seeded extractors in which the assumptions on the seed are relaxed. Namely, the~second, additional source of randomness (previously the seed) now only needs to have some known min-entropy and be independent of the initial output. Moreover, the~independence condition can also be relaxed---for~example, allowing coordination, cross-influence, or bounded mutual information with respect to the input~\cite{ball2022randomness} or independence only in the sense of a Markov chain~\cite{arnon2015quantum}\@.
  \item \textbf{{Level 4: Physical device-independent extractors}}---The last class that we consider are extractors requiring special additional hardware, providing the second randomness source needed in level 3 whilst making only minimal assumptions ({{for example}, that information cannot travel faster than the speed of light}\@). This is made possible by a particular type of interactive proof system in which quantum hardware can be verified to perform as promised, as~opposed to having to rely on modelling the physical process, as would be done normally. This `black box' verification gives a guaranteed lower bound on the min-entropy of the output, which can then be used together with the RNG's initial output in a two-source extractor as in level 3\@. These \textit{{physical}} extractors are referred to, in~the quantum information science community, as~device-independent randomness amplification protocols and have no classical analogue. 
  With today's technology, such extractors require making additional implementation assumptions (to the minimal ones)\@. 
  We return in detail to physical extractors in \Cref{subsec:rand-amp-processes-dirap}\@.
\end{itemize}

When a second bit string of randomness is required (levels 2 and 3), we use the NIST Randomness Beacon~\cite{kelsey2019reference}\@. 
For physical randomness extraction (level 4), we use a semi-device-independent randomness amplification protocol that is an adaptation of~\cite{foreman2020practical}, which we describe in \Cref{subsec:rand-amp-processes-dirap}\@. All the algorithms for extraction used in this work are from the software library \texttt{Cryptomite} \cite{cryptomite}, which can be found {at}  \url{https://github.com/CQCL/cryptomite}\@ (accessed on 28 November 2024). 

\begin{figure}[H]
	\includegraphics[width=0.75\linewidth]{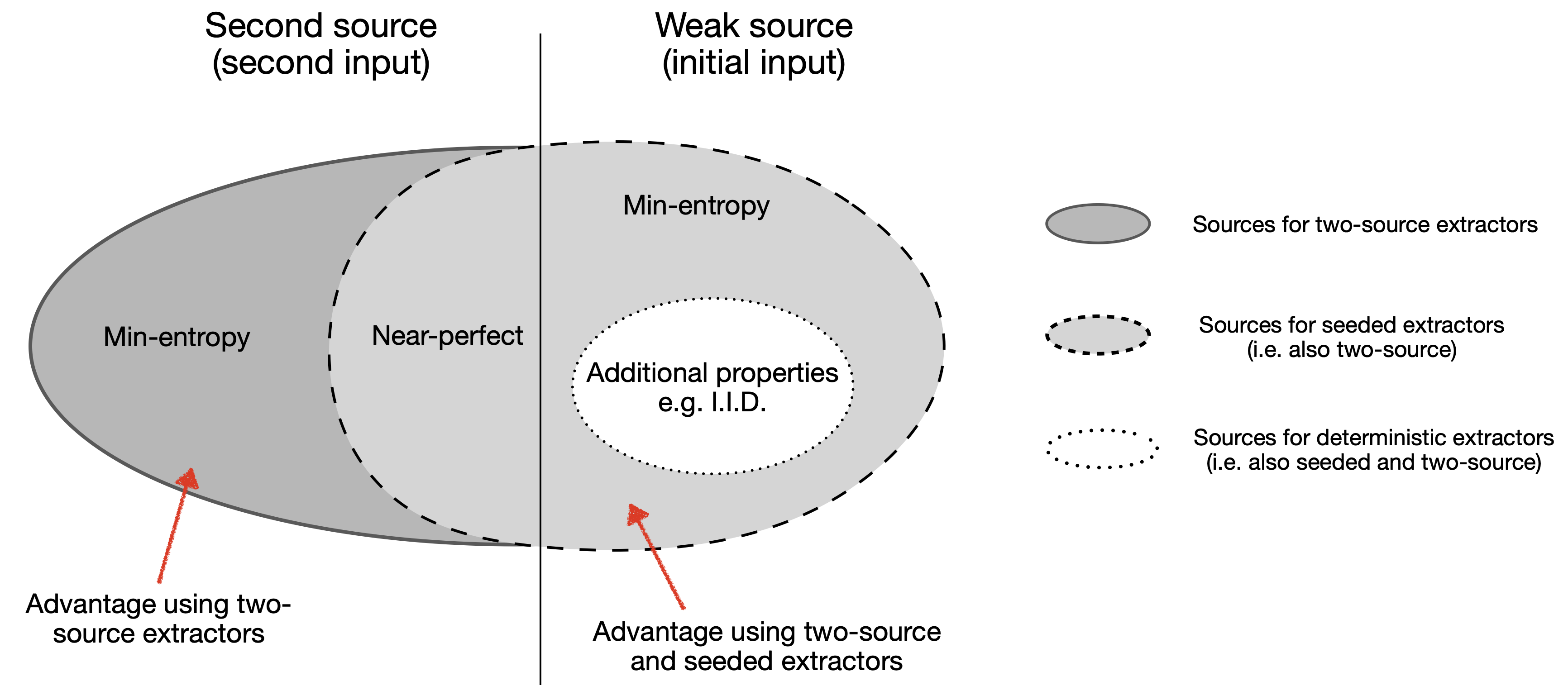}
	\caption{Illustration of the set of sources, or~input distributions, that can be successfully extracted from by different randomness extraction methods. (Right) weak input distributions and (Left) second input, or~weak seed, distributions. Deterministic extractors (level 1) require additional properties on the weak input but~do not need a second input source. Seeded extractors (level 2) relax the need for additional properties of the weak input and extract from sources with min-entropy only, at~the cost}
	\label{fig:rng-hierarchy}
\end{figure}

The assumptions that the different post-processing methods require are illustrated \mbox{in Figure~\ref{fig:rng-hierarchy}\@.}
\vspace{-6pt}
\vspace{-12pt}
{\captionof*{figure}{of requiring a second string of (near-)perfect randomness. Two-source extractors (level 3) relax the assumptions of seeded ones to a second source that also has min-entropy only. Physical extractors (level 4, not in the figure) require special quantum hardware, which effectively provides the second input with a device-independent lower bound on the min-entropy, requiring minimal added~assumptions.}}
\vspace{12pt}

\subsubsection{Results~Overview}
\label{subsubsec:results}

\begin{figure}[H]
  \includegraphics[width=0.48\textwidth]{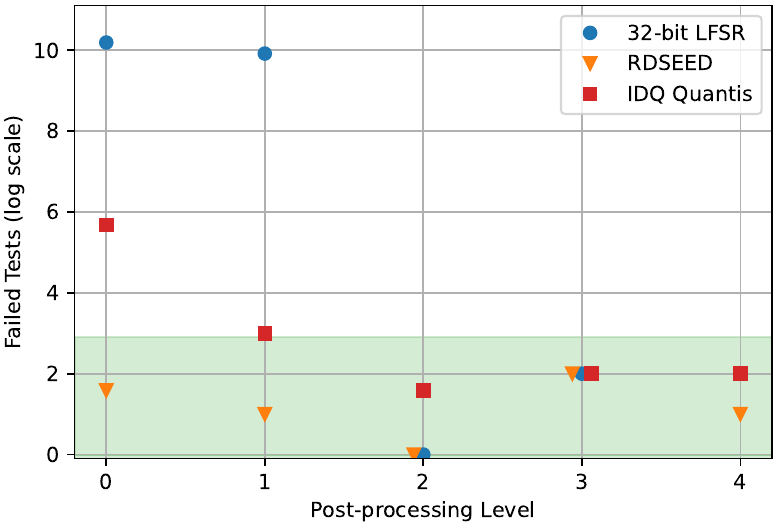}
  \includegraphics[width=0.48\textwidth]{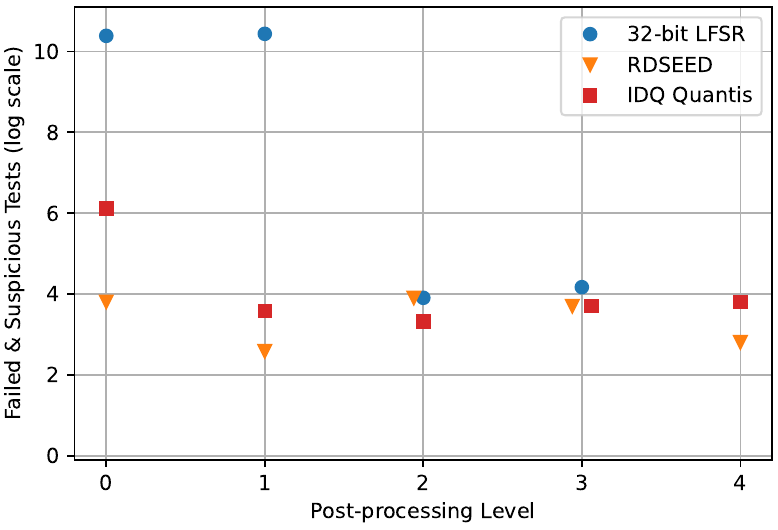}
  \caption{The above plots show (\textbf{left}) the number of statistical tests failed and (\textbf{right}) failed and suspicious for each initial RNG at each post-processing level. 
  The $x$ axis indicates the level, with~step 0 being the initial RNG with no additional post-processing, and steps 1--4 are deterministic, seeded, two-source, and physical extraction, respectively. 
  The $y$ axis is the number of statistical tests failed (\textbf{left}) or failed and suspicious (\textbf{right}), out of 4600, using a logarithmic scale: for $f$ failed or failed and suspicious tests, $y = \log_2(f+1)$. The~shaded region in the left plot illustrates the \textit{{successful}} region, whereby the RNG fails less than 7.5 tests, and~the white region illustrates the `unacceptable' region, in~which, with~high probability, near-perfect randomness is not produced. We note that we are unable to use the 32-bit LFSR at level 4 because of its low initial estimated min-entropy rate, $\alpha_{\mathsf{RNG}}$, as~detailed and evaluated in \Cref{sec:rng-analysis}\@.\label{fig:stats-1}}
\end{figure}

We now present the main results of the statistical testing of the different post-processing methods in Figure~\ref{fig:stats-1}, with~more details and tables in the following sections.
As stated before, we expect that an RNG producing near-perfect random numbers fails less than 7.5 of the 4600 tests that it is subject to, on~average, when testing $5 \times 10$ Gbit files. ({{This} number is the expected amount of type 2 error, i.e.,\ the expected maximum number of failed tests, given that the underlying distribution is indistinguishable from uniform. Note that we implicitly assume that each statistical test is independent}). This is the criterion that we use to call randomness generation \textit{{successful}} from a statistical perspective (green \mbox{highlighted area})\@.

Ideally, the~results would reflect the different levels of post-processing and the validity of the assumptions that these imply. 
Our results in Figure~\ref{fig:stats-1} tell a mixed story.
\setlist{nolistsep}
\begin{itemize}[noitemsep]
\item For the RNGs that fail the tests when unprocessed, we observe that additional post-processing indeed improves the quality of the initial output. Considering the LFSR, for~example, any extraction method higher than level 1 applied to the initial output produced a processed output that passed the numerical tests well.
IDQ's device, as~a second example, is significantly improved already with level 1 of extraction, but~only gives successful results when higher levels are applied.
\item Although level 3 is strictly a relaxation of the assumptions made at level 2, we were unable to observe a difference in the numerical results. 
This is because level 2, from~a statistical perspective, seems to be giving results that are already successful. 
Moreover, we are unable to distinguish between levels 2, 3, and 4. We interpret this as another illustration of the difference between statistical and cryptographic randomness, in~which weaker assumptions are desirable even if no statistical advantage can be witnessed from the user's perspective. 
It is also likely that, in~order to give non-trivial examples of step 2 failing, one would need to generate the seed in a manner that is either significantly biased or correlated to the RNG (both of which could happen in practice)\@.
\item All our implementations above level 1 gave successful numerical test results on the three RNGs that we tested. In~particular, from~a statistical perspective, this means that a poor PRNG (here, the 32-bit LFSR) can be concatenated with an extractor to form a good PRNG\@.
\end{itemize}

\subsection{Implementations of the Post-Processing~Methods}
\label{subsec:hierarchy_implementation}
We now describe how we implemented the post-processing, i.e.,\ different extractors in our levels, together with the parameter choices and compromises that we made. 
For the post-processing algorithms, we used the randomness extractors publicly available from the software library \Cryptomite~\cite{cryptomite}\@. 
{To assess the randomness quality at each step, we generated $5 \times 10$ Gbit test files of the processed output and performed statistical testing using the `all' setting (the most intense) in the $\STE$.} 
All randomness post-processing and statistical tests were run on a Dell Precision 7540 personal laptop with 16 GB of RAM and a 2.3 GHz Intel i9 processor, using the Ubuntu 20.04 operating system.
We state all input and output sizes and give detailed descriptions of each test setting and implementation of each level with the parameter choices, so that all results can be reproduced. For~each level, we chose the parameters of the different extractors such that, in~theory, the~processed output was $\epsilon_{\mathsf{total}}$-perfectly random (see \Cref{def:near-perfect-rand}), with~$\epsilon_{\mathsf{total}} \leq 2^{-32} \approx 10^{-10}$.

\subsubsection{Level 1: Deterministic~Extraction}
\label{subsec:rand-amp-processes-det-extraction}
A deterministic extractor will generate a near-perfectly random output when processing the initial output of RNGs with some well-defined properties. These well-defined properties vary depending on the extractor that is used, with~different choices~possible.

\begin{Definition}[Deterministic randomness extractor] \label{def:det-extractor}
  A deterministic randomness extractor is \mbox{a function}
\begin{align} \mathsf{Ext_d}: \{ 0,1\}^n \rightarrow \{0,1\}^m \end{align}
  such that, for~random variables $X \in \{ 0,1\}^n$ with \textit{specific properties} \cite{shaltiel2011introduction},
\begin{align} \Delta(\mathsf{Ext_d}(X), \mathrm{U}_m) \leq \epsilon\ , \end{align}
  where $\mathrm{U}_m$ is the uniform variable on $\{0,1\}^m$.
\end{Definition}

In other words, a~deterministic extractor is a function that maps random variables $X$ with specific characteristics to~a new variable $\mathsf{Ext_d}(X)$ that is near-perfectly random.
Note that the properties of $X$ required depend on the specific extractor---for example, that all bits in $X$ are I.I.D\@.

\begin{figure}[H]
  \includegraphics[width=0.75\textwidth]{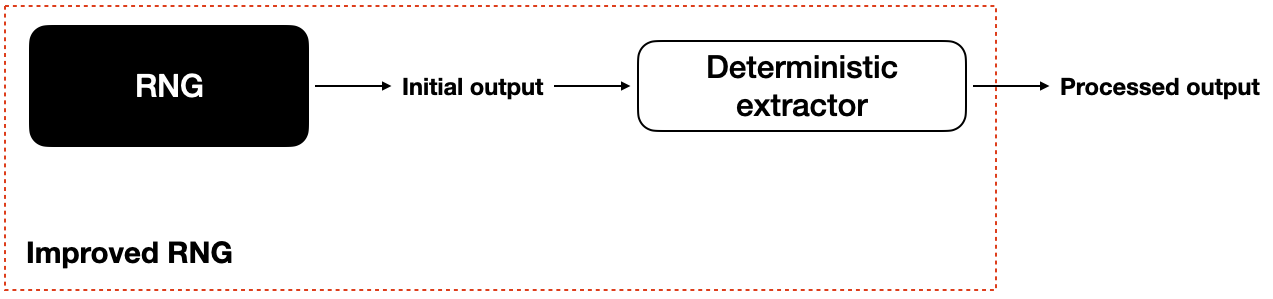}
  \caption{{Here,} 
 level 1 of our post-processing methods is performed by using a deterministic extractor, namely the \VonNeumann extractor, on~the initial output of the~RNG.}
  \label{fig:deterministic}
\end{figure}

The implementation of the deterministic extraction set-up is shown in Figure~\ref{fig:deterministic}\@. We use the \VonNeumann extractor~\cite{von1963various} to extract from the initial output $X \in \{0,1\}^n$ of the RNG, with~the implementation from~\cite{cryptomite}\@. 
This extractor requires that all two subsequent input bits have a fixed bias, i.e.,\ for bits $X_{2i}, X_{2i+1} \in \{0,1\}$ with $i= 1, \ldots, \lfloor \frac{n}{2} \rfloor$ and $p_i \in (0,1)$, we require that \begin{align} \label{eq:vn cond}
    \Pr(X_{2i} = 0) = \Pr(X_{2i+1} = 0) = p_i\ .
\end{align} 
The \VonNeumann extractor works by grouping subsequent bits in pairs and~outputting the first (or second) bit only when the bits in the pair are different, giving an output length of $m \approx p(1-p)$ (if the bias is fixed, $p_i=p$ for all $i$) and $\epsilon = 0$, i.e.,\ perfect randomness at the~output.

 The statistical test results in Table~\ref{table:vn-debiased-all} show the following.
\setlist{nolistsep}
\begin{itemize}[noitemsep]
\item Both the LFSR and IDQ Quantis show improvements compared to the initial RNG testing results (Table~\ref{table:all-tests-no-postprocessing}), although~they still do not pass overall.
\item The number of NIST statistical test failures for both the LFSR and IDQ Quantis increases when moving from no post-processing to deterministic post-processing. This may be due to specific biases in the RNGs that are incompatible with, or~even amplified by, the~\VonNeumann extractor (e.g., successive bits are not independent) or fundamental issues with the NIST tests, as~suggested in~\cite{burciu2019systematic,hamano2007correction,kowalska2022revision,marton2015interpretation}.
\end{itemize}

\begin{table}[H] 
\caption{{This} 
 table gives the sum of statistical tests failed for $5 \times 10$ Gbit samples from each RNG, after~deterministic extraction using the \VonNeumann extractor.
  Due to the 32-bit LFSR failing PractRand quickly, only 635 tests were conducted (instead of the full 4600), so we rescale these results. 
  In cells with multiple entries, failed tests are on the left and suspicious tests (when applicable) are on the right in parentheses. 
  Full results can be found in \Cref{app:vn-debiased}\@.\label{table:vn-debiased-all}}
  \begin{adjustwidth}{-\extralength}{0cm}
		\newcolumntype{C}{>{\centering\arraybackslash}X}
		\begin{tabularx}{\fulllength}{CCCCCCCC}
  \toprule
  \textbf{RNG}         & \begin{tabular}[c]{@{}c@{}}\textbf{NIST}\\ \textbf{(75)}\end{tabular} & \begin{tabular}[c]{@{}c@{}}\textbf{Diehard}\\ \textbf{(90)}\end{tabular} & \begin{tabular}[c]{@{}c@{}}\textbf{ENT}\\ \textbf{(30)}\end{tabular} & \begin{tabular}[c]{@{}c@{}}\textbf{SmallCrush}\\ \textbf{(75)}\end{tabular} & \begin{tabular}[c]{@{}c@{}}\textbf{Alphabit}\\ \textbf{(85)}\end{tabular} & \begin{tabular}[c]{@{}c@{}}\textbf{Rabbit}\\ \textbf{(200)}\end{tabular} & \begin{tabular}[c]{@{}c@{}}\textbf{PractRand}\\ \textbf{(4600)}\end{tabular} \\ \midrule
  32-bit LFSR & 25                                                    & 10  (5)                                                  & 5                                                  & 18                                                        & 76                                                       & 106                                                    & 724  (413)                                            \\ \midrule
  RDSEED      & 0                                                    & 0  (2)                                                   & 0                                                  & 0                                                         & 0                                                        & 1                                                      & 0  (2)                                                    \\ \midrule
  IDQ Quantis & 4                                                    & 0  (1)                                                   & 0                                                  & 0                                                         & 0                                                        & 3                                                      & 0  (3)                                                    \\ \midrule
  \end{tabularx}
 \end{adjustwidth}
\end{table}

Interestingly, applying the \VonNeumann extractor to an LFSR results in a stream cipher known as the self-shrinking generator, which has been studied for cryptographic use~\cite{meier1994self}. Although~the self-shrinking generator fails fewer tests compared to the unprocessed LFSR, a~substantial number of failures~remain.

{As noted at the start of this section, a~deterministic extractor can produce a near-perfectly random output if the input source satisfies certain specific properties. However, these properties are often difficult or even impossible to verify in practice. As~a result, it is more practical to base claims solely on the min-entropy of the source. In~\cite{santha1986generating}, it was demonstrated that deterministic extraction from a source characterised solely by min-entropy is impossible. Such sources require additional, independent randomness to \mbox{enable extraction}.}

\subsubsection{Level 2: Seeded~Extraction}
\label{subsec:rand-amp-processes-seed-extraction}
{Seeded extraction requires only a min-entropy guarantee for the initial RNG output but comes at the cost of needing a second, independent and~(near-)perfectly random input (the \textit{seed}) to enable extraction.}

\begin{Definition}[Seeded randomness extractor] \label{def:seeded-extractor}
  A seeded randomness extractor is a function 
  $\mathsf{Ext_s}: \{ 0,1\}^{n} \times \{0,1\}^{d} \rightarrow \{0,1\}^m$
  such that, for~a random variable $X \in \{ 0,1\}^{n}$ with min-entropy $\mathrm{H}_{\infty}(X) \geq k$, and~seed $S \in \{ 0,1\}^d$ with min-entropy $\mathrm{H}_{\infty}(S) =  d$ (i.e.,\ $S$ is perfectly~random),
\begin{align} \Delta(\mathsf{Ext_s}(X, S), \mathrm{U}_m) \leq \epsilon\ , \end{align}
  where $\mathrm{U}_m$ is the uniform distribution on $\{0,1\}^m$\@. 
\end{Definition}

A seeded extractor can be understood as a randomised function that maps a weakly random variable $X$ to a new variable $\mathsf{Ext_s}(X, S)$ that is (near-)perfectly random.
Note that the seed may be $\epsilon_s$-perfect only, with~an additive error in the statistical distance above, i.e.,\ $\epsilon \rightarrow \epsilon + \epsilon_s$ (see, for~example, Appendix A from~\cite{frauchiger4547true} for proof)\@. 
Seeded extractors are a special case of two-source extractors, which we define later in \Cref{def:2-extractor}.

\begin{figure}[H]
  \includegraphics[width=0.6\textwidth]{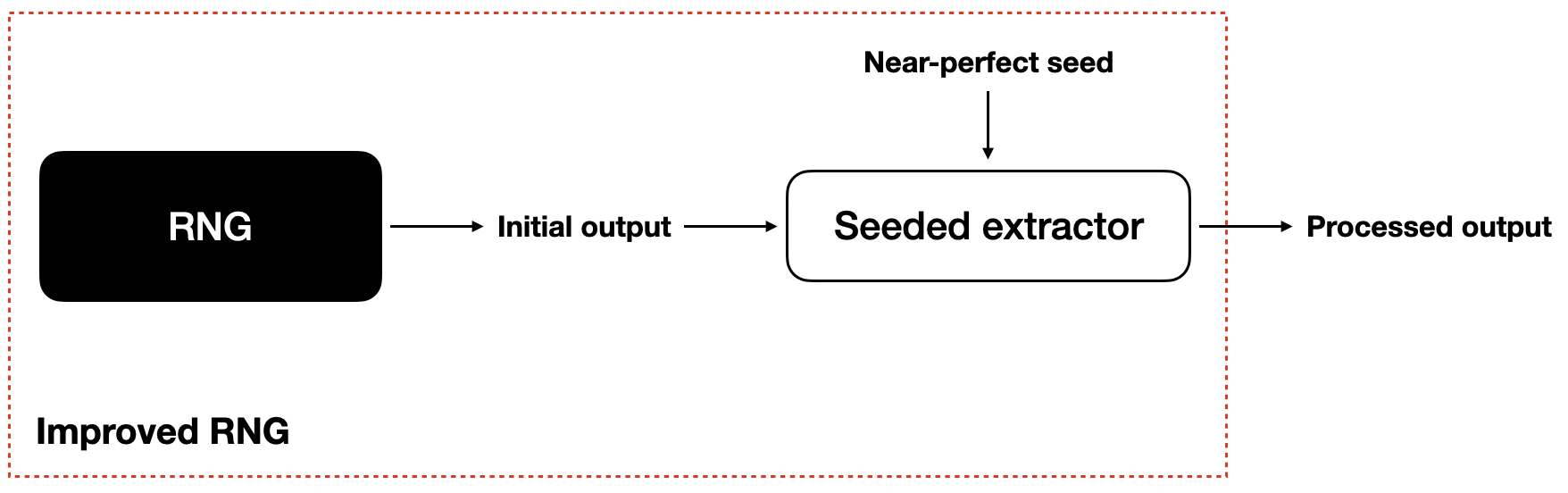}
  \caption{The set-up for seeded extraction. In~this case, the~initial output of the RNG only needs to have min-entropy, but~extraction requires an additional near-perfectly random bit string (the seed), which needs to be generated~independently.}
  \label{fig:seeded}
\end{figure}

\begin{Definition}[Strong seeded extractor] \label{def:strong-seeded-extractor}
  A \textit{strong} seeded randomness extractor is a function $\mathsf{Ext_s}: \{ 0,1\}^{n} \times \{0,1\}^{d} \rightarrow \{0,1\}^m$  such that
\begin{align}\Delta( [\mathsf{Ext_s}(X, S), S], [\mathrm{U}_m, S] ) \leq \epsilon\ ,\end{align}
  where $[\cdot,\cdot]$ denotes the concatenation of random variables and $\mathrm{U}_m$ is the uniform variable on $\{0,1\}^m$\@. 
\end{Definition} 
A strong seeded extractor is a randomised function that gives a (near-)uniform output, even when conditioned on the seed $S$ (the output is therefore independent of the seed)\@. 
This has some interesting consequences, which we exploit to generate the large amounts of processed output needed for statistical testing. 
Specifically, $S$ can be reused with different weak input random variables, allowing a single seed to be used in many extraction rounds.
The set-up for seeded extraction (implemented using a strong seeded extractor) is shown in Figure~\ref{fig:seeded}\@. 
The initial output from the RNG is split into blocks $X_i$ for $i = 1, \ldots, n$ with a promise on each block's min-entropy (\Cref{def:block-min-ent})\@. 

This step can be implemented with the \Circulant~\cite{cryptomite}, \Dodis~\cite{dodis2004improved}, \Toeplitz~\cite{krawczyk1994lfsr}, and~\Trevisan~\cite{trevisan1999construction} extractors from \Cryptomite, as~they can all be used as strong seeded extractors. 
Among these extractors, \Circulant offers the best trade-off between security parameters and computational complexity and~is therefore the one that we chose. 
The \Circulant extractor requires that the seed length is the input length plus one and~that the seed length is a prime. 
We set the seed length $|S|$ and RNG input block lengths \mbox{$|X_i|$ to $|S| = |X_i| + 1$ = 10,007}.
Note that using \Circulant allows us to generate cryptographic randomness even against an adversary able to store (and process) side information in quantum systems without changing the extraction algorithm, i.e.,~the extractor is \textit{{quantum-proof}}; see~\cite{cryptomite} for~details. 

To generate the seed $S$, we use the NIST Randomness Beacon, which is a public source of randomness produced by the US Government agency (NIST), mixing different randomness sources together, including chaotic classical and quantum processes~\cite{kelsey2019reference}\@. 
The min-entropy $k_i^{\mathsf{RNG}}$ for each block $X_i$ is $k_i^{\mathsf{RNG}} = \alpha_{\mathsf{RNG}}|X_i|$, where $\alpha_{\mathsf{RNG}}$ is a lower bound on the min-entropy per bit for each initial RNG block of outputs $X_i$, with~probability $\epsilon_{\mathsf{est}} < 2^{-32}$ (as found in \Cref{eps_est} {of} 
\Cref{app:min-ent-bound})\@. 
The output length after extraction, $m$, is then roughly $m \approx k_i^{\mathsf{RNG}}$\@. 

In order to generate the required $5 \times 10$ Gbits of processed output, the~\Circulant extractor is used multiple times on different initial output blocks $X_i$ with the same seed.
The extractor's outputs are then concatenated together until a final output, $\mathsf{Output}$, of~sufficient size is generated. 
The $\mathsf{Output}$ is given by
\begin{align} \mathsf{Output} = \left[\mathsf{Ext_s^{Circulant}}(X_1,S),\mathsf{Ext_s^{Circulant}}(X_2,S), \ldots, \mathsf{Ext_s^{Circulant}}(X_n,S) \right]\ , \end{align} where $[\cdot,\cdot]$ denotes the concatenation of random variables.
Each extraction round, which we index $i$, has an associated error $\epsilon_{\mathsf{ext_i}}$, and we choose the total security parameter to be $\epsilon_{\mathsf{total}}\leq 2^{-32}$---namely, everything is chosen so that $\epsilon_{\mathsf{total}} = \epsilon_{\mathsf{est}} + \sum_{j=1}^n \epsilon_{\mathsf{ext_j}}\leq 2^{-32}$. 
This derivation for $\epsilon_{\mathsf{total}}$, specifically that the composed output error is the sum of each of the individual extractor errors, can be found in~\cite{cryptomite}\@. 

\noindent The observations that we draw from the results in Table~\ref{table:nist-seeded-ext} are the following.
\setlist{nolistsep}
\begin{itemize}[noitemsep]
\item The statistical test results show a significant improvement in the results using deterministic extraction; see \Cref{subsec:rand-amp-processes-det-extraction}\@. In~particular, all RNGs have been successfully post-processed from a statistical perspective.
\item Even the 32-bit LFSR is successfully extracted from, which suggests that one can, from~a statistical perspective, build good PRNGs by appending an extractor to \mbox{poor PRNGs}.
\item Randomness that has a small amount of min-entropy only can pass statistical tests successfully. 
This is somewhat unsurprising as cryptographically secure PRNGs exist, but~we find it interesting to comment on nonetheless. 
The total entropy of the final output of the processed LFSR output is upper-bounded by 10,007 $+32$ (the seed length of the extractor plus the seed length of the 32-bit LFSR), in~the 50 Gbit of processed output generated, i.e.,\ a true min-entropy rate of, at~most, \mbox{$\alpha = (10,007 + 32)/ (5 \times 10^{10}) <10^{-5}$.}
\end{itemize}

\begin{table}[H] 
\caption{{This} 
 table gives the sum of statistical tests failed for $5 \times 10$ Gbit samples from each RNG, after~a strong seeded extractor has been applied to its initial output. The~seed is generated using the NIST Randomness Beacon.
  In cells with multiple entries, failed tests are on the left and suspicious tests (when applicable) are on the right. 
  Full results can be found in \Cref{app:nist-se}\@.} \label{table:nist-seeded-ext}
  \begin{adjustwidth}{-\extralength}{0cm}
		\newcolumntype{C}{>{\centering\arraybackslash}X}
		\begin{tabularx}{\fulllength}{CCCCCCCC}
  \toprule
  \textbf{RNG}         & \begin{tabular}[c]{@{}c@{}}\textbf{NIST}\\ \textbf{(75)}\end{tabular} & \begin{tabular}[c]{@{}c@{}}\textbf{Diehard}\\ \textbf{(90)}\end{tabular} & \begin{tabular}[c]{@{}c@{}}\textbf{ENT}\\ \textbf{(30)}\end{tabular} & \begin{tabular}[c]{@{}c@{}}\textbf{SmallCrush}\\ \textbf{(75)}\end{tabular} & \begin{tabular}[c]{@{}c@{}}\textbf{Alphabit}\\ \textbf{(85)}\end{tabular} & \begin{tabular}[c]{@{}c@{}}\textbf{Rabbit}\\ \textbf{(200)}\end{tabular} & \begin{tabular}[c]{@{}c@{}}\textbf{PractRand}\\ \textbf{(4600)}\end{tabular} \\ \midrule
  32-bit LFSR & 0                                                    & 0  (3)                                                   & 0                                                  & 0                                                         & 0                                                        & 0                                                      & 0  (6)                                                    \\ \midrule
  RDSEED      & 0                                                    & 0  (7)                                                   & 0                                                  & 0                                                         & 0                                                        & 0                                                      & 0  (7)                                                    \\ \midrule
  IDQ Quantis & 0                                                    & 0  (2)                                                   & 0                                                  & 0                                                         & 0                                                        & 2                                                      & 0  (5)                                                    \\ \bottomrule
  \end{tabularx}
 \end{adjustwidth}
\end{table}

Our results at this level are disappointing, in~the sense that the successful test results mean that we will not be able to distinguish the next levels (3 and 4) from level 2 from a statistical perspective---for example, that level 3 is strictly better than level 2\@. 
It would be interesting to find non-trivial examples where the output of a seeded extractor fails statistical tests because of a seed generated in a way that is not independent or near-uniform. 
Unfortunately, we could only find artificial examples (i.e.,\ when all seed bits are the same) that were detected by our statistical testing. 

\subsubsection{Level 3: Two-Source~Extraction}
\label{subsec:rand-amp-processes-2-source-extraction}
Seeded extraction (level 2) requires an independent string of (near-)perfect randomness as an initial resource, which is difficult to justify and leads to circularity: one needs near-perfect randomness to generate more of it. 
Two-source extraction relaxes this requirement, allowing the second string to be only weakly random, in~the sense that it has some min-entropy and/or 
only a relaxed notion of independence ({{for example}, the~case of using a two-source extractor secure in the Markov model~\cite{arnon2015quantum}, where the two input sources can be correlated through a common cause, or~if the sources may have bounded coordination, cross-influence, or mutual information~\cite{ball2022randomness}\@})---although, in this work, we calculate our two-source extractor parameters based on standard independence between the two input sources. 
Two-source extractors can be used as seeded extractors, simply by assuming that one of the input strings is already near-perfect and independent; therefore, level 3 is strictly a relaxation of the assumptions of level 2\@.

\begin{Definition}[Two-source randomness extractor] \label{def:2-extractor}
  A two-source randomness extractor is a function $\mathsf{Ext_2}: \{ 0,1\}^{n_1} \times \{0,1\}^{n_2} \rightarrow \{0,1\}^m $
  such that, for~statistically independent random variables $X \in \{ 0,1\}^{n_1}$ and $Y \in \{ 0,1\}^{n_2}$ with min-entropy $\mathrm{H}_{\infty}(X) \geq k_1$ and\linebreak   $\mathrm{H}_{\infty}(Y) \geq k_2$, respectively,
\begin{align} \Delta(\mathsf{Ext_2}(X, Y), \mathrm{U}_m) \leq \epsilon\ , \end{align}
  where $\mathrm{U}_m$ is the uniform variable on $\{0,1\}^m$. 
\end{Definition}
In other words, a~two-source extractor is a weakly randomised function that maps a random variable $X$ to a new variable $\mathsf{Ext_2}(X, Y)$ that is~near-perfect.

\begin{figure}[H]
  \includegraphics[width=0.75\textwidth]{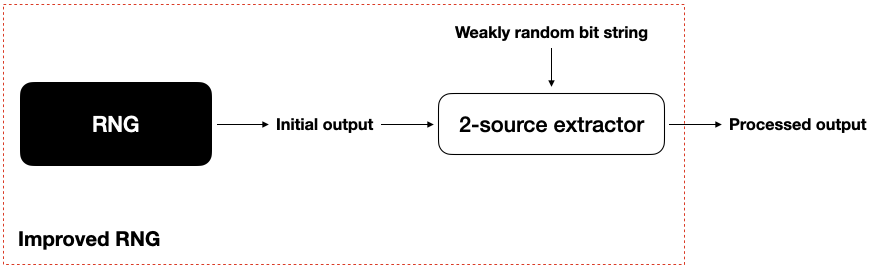}
  \caption{The set-up for two-source extraction. In~this case, the~initial output of the RNG only needs to have some min-entropy and extraction requires an additional bit string that is weakly random only in~the sense that it also has~min-entropy.}
  \label{fig:two-source}
\end{figure}

\begin{Definition}[Strong two-source extractor] \label{def:strong-2-extractor}
  A two-source randomness extractor is said to be \textit{strong} in the input $Y$ if the function $\mathsf{Ext_2}$ is such that
\begin{align} \Delta([\mathsf{Ext_2}(X, Y), Y], [\mathrm{U}_m Y] ) \leq \epsilon\ , \end{align}
  where $[\cdot,\cdot]$ denotes the concatenation of random variables and $\mathrm{U}_m$ is the uniform variable on $\{0,1\}^m$\@.
\end{Definition}
Strong two-source extractors, like strong seeded extractors, allow for one input source to be used in multiple extraction rounds. The~set-up for seeded extraction (implemented using a strong seeded extractor) is shown in Figure~\ref{fig:two-source}\@. 
\vspace{-6pt}
From the $\Cryptomite$ library, we again use the \Circulant extractor~\cite{cryptomite}, but~this time as~a strong two-source extractor. 
Other extractors in $\Cryptomite$ can be used too, but, since the \Circulant extractor offers the best parameters and efficiency, we use it in our implementation. For~full details, we refer the reader to~\cite{cryptomite}\@.
Two-source extraction requires a second input source with min-entropy above some threshold based on the specific two-source extractor construction.
For the \Circulant extractor, this requirement is that the sum of the min-entropy rates of the two weak inputs is at least 1\@.
$X_i$ is the initial RNG output blocks and $Y$ is the additional weakly random input (which we sometimes call the \textit{{weak seed}}) and, as~in level 2, we set $|Y| = |X_i| + 1 =$ {10,007}
\@.

To generate $Y$, we again use the NIST Randomness Beacon, but, in~this case, we minimise the amount of entropy that we assume that it contains, instead of assuming that it has full entropy as in level 2\@. 
This change in the assumption increases the likelihood that the assumption holds in practice.
The output length of the \Circulant extractor is roughly $(\alpha_{\mathsf{NIST}} + \alpha_{\mathsf{RNG}} - 1) |Y|$, which we impose by adjusting the min-entropy rate assumption of the NIST Randomness Beacon as $\alpha_{\mathsf{NIST}}$, as~\begin{align} 
    \alpha_{\mathsf{NIST}} = 1.02 - \alpha_{\mathsf{RNG}}\,
\end{align} where $\alpha_{\mathsf{RNG}}$ is the min-entropy rate of the initial RNG (found in \Cref{sec:rng-analysis})\@. 
We use 1.02 instead of 1 to account for spurious terms in the parameter calculation that reduce the output length; see~\cite{cryptomite} for the explicit calculation of these penalty terms. 
In other words, we use the computed min-entropy rate of the RNG under study to minimise the assumption about the second source's min-entropy rate, whilst imposing a non-trivial output length from the~extractor.

The processed output is then generated in two steps: (1) using the \Circulant extractor as a two-source extractor on the two input strings $X_1$ and $Y$, we generate a (near-)perfect output, which will be the seed in the next step; (2) we use this seed in multiple \Circulant seeded extractions on $X_{i\geq 2}$. The~multiple outputs of the seeded extractor are concatenated together to obtain a final output of $5 \times 10^{10}$ bits. 
In other words, the~concatenation of the two-source and seeded extractors together forms a two-source extractor with advantageous parameters. Therefore, the~final output for statistical testing is given by
\begin{align} \label{eq:out2} 
\mathsf{Output} &= \left[\mathsf{Ext_s^{Circulant}}(X_2,S),\mathsf{Ext_s^{Circulant}}(X_3,S), \ldots, \mathsf{Ext_s^{Circulant}}(X_n,S) \right]\ ,
\end{align}
where $S = \mathsf{Ext_2^{Circulant}}(X_1, Y)$, $[\cdot,\cdot]$ denotes the concatenation of random variables and the extractor round with input $X_i$ has error $\epsilon_{\mathsf{ext_i}}$. 
The total error of the final output is $\epsilon_{\mathsf{total}} = \epsilon_{\mathsf{est}} + \epsilon_{\mathsf{ext_1}} + \sum_{j=2}^n \epsilon_{\mathsf{ext_j}} \leq 2^{-32}$. 
Proof that a strong two-source extractor and strong seeded extractor can be composed into $\mathsf{Ext_s^{Circulant}}(X_{i > 1},S)$, for~$S$ the output of a two-source extractor (right-hand side of \Cref{eq:out2}) can be found in~\cite{vadhan2012pseudorandomness} {Section~6.3}.
This, combined with the fact that the composed output error is the sum of each of the individual extractor errors (in~\cite{cryptomite}), allows us to calculate $\epsilon_{\mathsf{total}}$\@.

Our results in Table~\ref{table:nist-2-ext} show that all RNGs extracted at level 3 are successful from a statistical perspective, as in the seeded extraction case (level 2).
In the Appendices, we implement a variant of level 3 (two-source extraction) where all input strings are drawn from the initial RNG and there is no randomness from an alternative RNG, i.e.,\ rewriting the $\mathsf{Output}$ in \Cref{eq:out2} using $Y=X_0$, where $X_0$ is another output block from the initial RNG\@. 
In this regime, for~near-perfect randomness to be generated, all blocks produced by the initial RNG must be independent of one another (as well as having block min-entropy)\@. 
Even in this case, the~results were successful statistically. 
Full explanations and results can be found in \Cref{app:2-source-independence}\@.
\vspace{-6pt}
\begin{table}[H]
  \caption{{This} 
 table gives the sum of statistical tests failed for $5 \times 10$ Gbit samples from each RNG, after~strong two-source extraction, taking the RNG as one weak source and randomness from the NIST Randomness Beacon as the second.
  In cells with multiple entries, failed tests are on the left and suspicious tests (when applicable) are on the right in parentheses. 
  Full results can be found in \Cref{app:nist-2e}\@.} \label{table:nist-2-ext}
  \begin{adjustwidth}{-\extralength}{0cm}
		\newcolumntype{C}{>{\centering\arraybackslash}X}
		\begin{tabularx}{\fulllength}{CCCCCCCC}
  \toprule
  \textbf{RNG}         & \begin{tabular}[c]{@{}c@{}}\textbf{NIST}\\ \textbf{(75)}\end{tabular} & \begin{tabular}[c]{@{}c@{}}\textbf{Diehard}\\ \textbf{(90)}\end{tabular} & \begin{tabular}[c]{@{}c@{}}\textbf{ENT}\\ \textbf{(30)}\end{tabular} & \begin{tabular}[c]{@{}c@{}}\textbf{SmallCrush}\\ \textbf{(75)}\end{tabular} & \begin{tabular}[c]{@{}c@{}}\textbf{Alphabit}\\ \textbf{(85)}\end{tabular} & \begin{tabular}[c]{@{}c@{}}\textbf{Rabbit}\\ \textbf{(200)}\end{tabular} & \begin{tabular}[c]{@{}c@{}}\textbf{PractRand}\\ \textbf{(4600)}\end{tabular} \\  \midrule
  32-bit LFSR & 0                                                  & 0  (6)                                                   & 0                                                  & 0                                                         & 2                                                        & 1                                                      & 0  (8)                                                    \\  \midrule
  RDSEED      & 0                                                  & 0  (4)                                                   & 0                                                  & 0                                                         & 0                                                        & 3                                                      & 0  (5)                                                    \\ \midrule
  IDQ Quantis & 0                                                   & 0  (3)                                                   & 0                                                  & 0                                                         & 0                                                        & 1                                                      & 0  (5)                                                    \\ \bottomrule
  \end{tabularx}
  \end{adjustwidth}
\end{table}

\subsubsection{Level 4: Physical Randomness~Extraction}
\label{subsec:rand-amp-processes-dirap}
Two-source extraction (level 3) allows for the generation of near-perfect randomness if two weakly random but independent strings of randomness are available. 
In the final level, we consider post-processing with a \textit{{physical}} randomness extractor. 
This level is called physical because it requires a quantum device, in~addition to the initial RNG, while the other levels only require mathematical algorithms to perform extraction. 
At a high level, the~role of this additional hardware is to provide a second string of random numbers, whilst making minimal assumptions~only.

\begin{figure}[H]
  \includegraphics[width=0.75\textwidth]{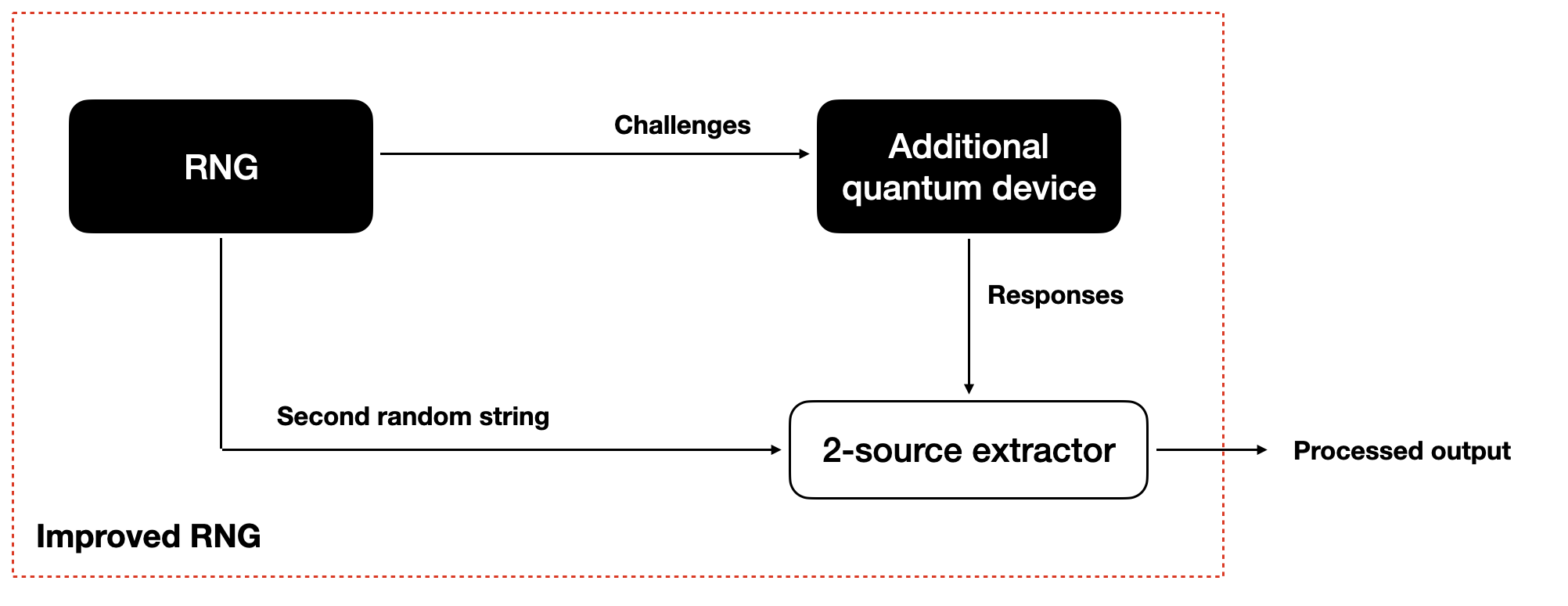}
  \caption{The set-up for level 4: physical randomness extraction. The~initial RNG is used twice: first to generate challenges to the quantum device and~second to~provide an extra bit string as input to a two-source extractor. The~role of the quantum device is to provide an additional source of randomness. The~device-independent protocol is performed by using the challenge--response behaviour of the device to obtain a lower bound on the amount of randomness in the device's responses (without characterising the device itself)\@. The~second bit string of the initial RNG and the responses from the quantum device form the two input strings to a two-source extractor, implemented as in level 3\@.}
  \label{fig:physical-ext}
\end{figure}

Adding quantum hardware may initially seem to imply introducing numerous assumptions; however, following the \textit{{device-independent}} approach, this hardware can, in principle, be treated as an untrusted \textit{{black box}} (which could even have been built by an adversary, so long as it can be shielded once in use and meets some minimal requirements). 
We call the added assumptions \textit{{minimal}} because they are either fundamental to physics---e.g.,\ information cannot travel faster than light speed---or no cryptography can ever be performed without them---e.g.,\ the devices are shielded (there are no backdoors)\@. 
This is made possible by the development of device-independent protocols, which rely on Bell tests~\cite{brunner2014bell}\@. 
The idea is to use the initial RNG to generate random challenges for the quantum device and~then study its response. With~ideal (noiseless) devices, this approach can be used to \textit{{self-test}} the inner functioning of the device, i.e.,\ one can uniquely identify the implemented quantum states and measurements from the observed challenge--response statistics alone. For~real (noisy) devices, this approach can be used to bound the adversary's guessing power, and~thus guarantee min-entropy, over~the device's outputs or responses.
This approach crucially relies on quantum resources, which have this self-testing property, and~has no classical analogue. 
For a review on the subject, together with its minimal assumptions (called loopholes), we refer the reader to~\cite{acin2016certified}\@. 
See Figure~\ref{fig:physical-ext} for an~illustration.

Today, quantum devices that are capable of running device-independent protocols are extremely difficult to build (they require the ability to perform a loophole-free Bell Test~\cite{brunner2014bell}) and exist as experiments on lab benches only. 
Because of this, more practical implementations have been developed in which a few well-justified assumptions are added (to the minimal ones)\@. The~resulting protocols have comparatively fewer assumptions than standard hardware, but~not fundamentally minimal. 
Such a \textit{{semi-}}device-independent protocol is the one that we implement for our physical extraction method at level 4, based on an adaptation of the randomness amplification protocol described in~\cite{foreman2020practical} and implemented on quantum computers. For~clarity, the~assumptions that we make are the following.
\setlist{nolistsep}
\begin{itemize}[noitemsep]
  \item The initial RNG has a block min-entropy structure (as in seeded and two-source extraction)\@. 
  \item The quantum device is independent of the initial RNG's output; we do not consider correlations between the two (although this can be added)\@. This assumption is well motivated since the quantum computer is distant from the initial RNG. 
  \item The quantum device is assumed to perform a faithful Bell test. This assumption is well motivated when using particular types of devices, such as the quantum computers based on ion traps that we use---see the discussion in~\cite{foreman2020practical} ({Section 6.2}
, \textit{{Validity of quantum computers for Bell experiments and added assumptions}})\@.
\end{itemize}

We used the $\mathsf{H}$1-1 Quantinuum ion-trap quantum computer~\cite{h1-1} as our device to obtain, from~its output, a~weakly random bit string size of $3.6 \times 10 ^ 6$ bits ({{this means that}, due to using the \Circulant extractor, the~input length of the initial RNG block to the two-source extraction is also $3.6 \times 10^6$ bits}) with min-entropy rate $\alpha_{\mathsf{Q}} \geq 0.518$, certified in the semi-device-independent manner described above. 
The \Circulant extractor requires $\alpha_{\mathsf{Q}} + \alpha_{\mathsf{RNG}} > 1$ to give a non-vanishing output, implying that the rate of an initial RNG must satisfy $\alpha_{\mathsf{RNG}} > 0.482$ to allow for physical extraction with our implementation. ({{This minimum} requirement is particularly interesting, since, even if one has access to two identical (and independent) copies of an initial RNG with $\alpha_{\mathsf{RNG}} = 0.482 + \delta$ for $\delta \in (0, 0.18)$, one would be unable to extract from the two (step 3) with today's implemented extractors. Note that this is not a fundamental limitation, as~other two-source extractors allow for one of the strings to have a logarithmic min-entropy rate only. However, ~to our knowledge no such extractor has been implemented, let alone efficiently}\@). The~advantage of using a quantum device, and~therefore level 4, is two-fold: (a) one obtains a rigorous, semi-device-independent lower bound on a second bit string's min-entropy and (b) the min-entropy rate of the quantum device is above $0.5$, allowing extraction from a weak initial output with rate $0.5$ using the \Circulant extractor. Note that the min-entropy of the LSFR was too low to perform physical extraction (its min-entropy rate was below $0.482$; see Table~\ref{table:min-ent-est})\@.

The processed output is then generated in two steps. (1) We generate a (near-)perfect seed using the \Circulant extractor as a two-source extractor on the two input strings $X_1$, from~the initial RNG, and~$Y$, from~the $\mathsf{H}$1-1 Quantinuum quantum computer. (2) We use this seed in multiple \Circulant seeded extractions on $X_{i\geq 2}$, which are concatenated together to obtain a final output of $5 \times 10^{10}$ bits. 
In other words, the~concatenation of the two-source and seeded extractors together again forms a two-source extractor with advantageous parameters. 
Therefore, the~final output for statistical testing is given by
\begin{align} \mathsf{Output} =\left[\mathsf{Ext_s^{Circulant}}(X_2,S),\mathsf{Ext_s^{Circulant}}(X_3,S), \ldots, \mathsf{Ext_s^{Circulant}}(X_n,S) \right]\ ,
\end{align} where $S = \mathsf{Ext_2^{Circulant}}(X_1, Y)$, $[\cdot,\cdot]$ denotes concatenation and the extractor round with input $X_i$ has error $\epsilon_{\mathsf{ext_i}}$. The~total error of the final output is $\epsilon_{\mathsf{total}} = \epsilon_{\mathsf{est}} + \epsilon_{\mathsf{ext_1}} + \sum_{j=2}^n \epsilon_{\mathsf{ext_j}} \leq 2^{-32}$. This last step is similar to that of level 3, where the NIST Randomness Beacon is replaced by the $\mathsf{H}$1-1 Quantinuum quantum computer. The~statistical test results are given in Table~\ref{table:sdi-ext}.

\begin{table}[H]
  \caption{This table gives the sum of statistical tests failed for $5 \times 10$ Gbit samples from level 4\@.
  Note: The 32-bit LFSR does not generate any output in this setting, since its min-entropy is too low for extraction. 
  In cells with multiple entries, failed tests are on the left and suspicious tests (when applicable) are on the right in parentheses. 
  Full results can be found in \Cref{app:sdi}\@.} \label{table:sdi-ext}
\begin{adjustwidth}{-\extralength}{0cm}
 \newcolumntype{C}{>{\centering\arraybackslash}X}
\begin{tabularx}{\fulllength}{CCCCCCCC}
  \toprule
  \textbf{RNG}         & \begin{tabular}[c]{@{}c@{}}\textbf{NIST}\\ \textbf{(75)}\end{tabular} & \begin{tabular}[c]{@{}c@{}}\textbf{Diehard}\\ \textbf{(90)}\end{tabular} & \begin{tabular}[c]{@{}c@{}}\textbf{ENT}\\ \textbf{(30)}\end{tabular} & \begin{tabular}[c]{@{}c@{}}\textbf{SmallCrush}\\ \textbf{(75)}\end{tabular} & \begin{tabular}[c]{@{}c@{}}\textbf{Alphabit}\\ \textbf{(85)}\end{tabular} & \begin{tabular}[c]{@{}c@{}}\textbf{Rabbit}\\ \textbf{(200)}\end{tabular} & \begin{tabular}[c]{@{}c@{}}\textbf{PractRand}\\ \textbf{(4600)}\end{tabular} \\ \midrule
  32-bit LFSR & -                                                   & -                                                      & -                                                  & -                                                         & -                                                        & -                                                      & -                                                          \\ \midrule
  RDSEED      & 0                                                  & 0  (2)                                                   & 0                                                  & 0                                                         & 0                                                        & 1                                                      & 0  (3)                                                    \\ \midrule
  IDQ Quantis & 0                                                   & 0  (3)                                                   & 1                                                  & 0                                                         & 0                                                        & 2                                                      & 0  (7)                                                    \\ \bottomrule
  \end{tabularx}
  \end{adjustwidth}
\end{table}

The results of the statistical tests in Table~\ref{table:sdi-ext} show, as~for levels 2 and 3, that the post-processed RNGs perform well at level~4.

\section{Discussion}
In this work, we have presented a variety of extraction methods to post-process the output of random number generators (RNGs) and evaluated their impacts on the statistical properties of three widely used RNGs. We started by extensively testing the output from three RNGs and identified statistical failures in two of them, corroborating and extending previous findings~\cite{hurley2020quantum, hamburg2012analysis}. For~the RNGs that failed, all post-processing methods improved the statistical properties. Specifically, we found that the processed outputs processed with level 2 or higher (seeded, two-source, or~physical extraction) were statistically indistinguishable from uniform distributions. However, due to the inherent limitations of statistical testing, we could not identify any examples that failed under level 2 post-processing (seeded extraction) but succeeded with level 3 or higher (two-source and physical extraction), even though level 3 is provably stronger than level~2.

Our statistical testing environment, $\STE$, software, documentation, and build file can be found {at} \url{https://github.com/CQCL/random\_test} (accessed on 28 November 2024), and~the randomness extractor software library \Cryptomite can be found {at} \url{https://github.com/CQCL/cryptomite} (accessed on 28 November 2024) and in~\cite{cryptomite}. These tools may be independently interesting, making both statistical testing and randomness extraction easy to use and openly~accessible.  

A number of interesting future directions arise. It would be interesting to perform the statistical testing of other RNGs with our test environment to analyse how they perform when tested beyond what is required by standardisation bodies. Similarly, it would be interesting to include different post-processing methods than the ones that we have presented. One could use, for~example, vetted conditioning components from NIST~\cite{turan2018recommendation} and compare their results to the ones obtained using information-theoretic randomness~extractors.

We could have moved even further in our numerical testing but, because~numerical tests consume substantial computational resources, decided to omit certain test suites from our analysis, including SPRNG~\cite{mascagni2000algorithm} and Crypt-X~\cite{gustafson1994computer}.
Moreover, we were recently made aware of the numerical tests BitReps~\cite{uokbitreps} and RaBiGeTe~\cite{rabi}, which are also not included in $\STE$\@.
It would be interesting to include these in the analysis to obtain an even more intense statistical testing~environment.

\vspace{6pt}

\authorcontributions{Conceptualisation, C.F. and F.J.C.; methodology, C.F. and F.J.C.; software, R.Y.; validation, C.F., R.Y. and F.J.C.; formal analysis, C.F.; investigation, C.F. and F.J.C.; resources, F.J.C.; data curation, C.F.; writing—original draft preparation, C.F. and F.J.C.; writing—review and editing, C.F., R.Y. and F.J.C.; visualisation, C.F., R.Y. and F.J.C.; supervision, F.J.C.; project administration, C.F. and F.J.C. All authors have read and agreed to the published version of the manuscript.}

\funding{This research received no external~funding.}

\institutionalreview{Not applicable.}

\dataavailability{The data presented in this study are openly available in \url{https://github.com/CQCL/random\_test} (accessed on 28 November 2024). 
} 

\acknowledgments{We thank Erik Woodhead and Ela Lee for the useful discussions and~suggestions.}

\conflictsofinterest{The authors declare no conflicts of~interest.} 

\appendixtitles{yes} 
\appendixstart
\appendix

\section{RNG~Descriptions}
\label{app:rng-defs}
\unskip

\subsection{Linear Feedback Shift Register (LFSR)}
\label{app:lfsr}
The LFSR is a class of pseudo-RNGs that is commonly used in applications due to its speed and ease of implementation in both software and hardware, e.g.,\ \cite{datta2017design,sahithi2012implementation}\@.
Notably, LFSRs are used in cryptography, including in hashing and authentication~\cite{krawczyk1994lfsr} and stream ciphers~\cite{ekdahl2003lfsr}\@. 
For this work, we implement the maximal period LFSR found in~\cite{panda2012fpga}\@.

This LFSR generates pseudo-randomness as follows. Let $s = b_1, \ldots, b_{32}$ denote the initial 32-bit state, where $b_i$ denotes the $i=1, \ldots, 32$th bit. 
\setlist{nolistsep}
\begin{enumerate}[noitemsep]
  \item Initialise the LFSR with the 32-bit initial state $s = b_1, b_2, ... , b_{31}, b_{32}$.
  \item Calculate the feedback $f$ of $s$, where $f = b_{32} \oplus b_{22} \oplus b_{2} \oplus b_{1} \oplus 1$, where $\oplus$ denotes addition modulo 2.
  \item Output bit $b_1$.
  \item Replace bit $b_i$ with bit $b_{i-1}$ for all $i \in (2, 32)$.
  \item Set $b_{32} = f$. 
  \item Repeat steps (2--5) until the desired amount of bits has been generated. 
\end{enumerate} The maximum period for a 32-bit LFSR is $2^{32} - 1$. This means that bits repeat every $2^{32} - 1$ generated bits (approximately every $4.3$ Gbits)\@.

\subsection{Intel~RDSEED}
\label{app:thermal-diode}
Intel manufactures a hardware RNG based on thermal noise, which is present in their computer processing units. 
This true-RNG is constructed as follows, although~a more in-depth description can be found in~\cite{jun1999intel, hamburg2012analysis}\@.
\setlist{nolistsep}
\begin{enumerate}[noitemsep]
  \item Initial weak randomness is generated from an entropy source. This source is a self-clocking circuit designed such that, when the clock is running, the~circuit enters a meta-stable state, which then resolves to one of two possible states---determined randomly by thermal noise. 
  The state in which the circuit resolves is the random bit output from the entropy source. 
  This self-clocking occurs irregularly at around 3 GHz.
  \item Health and swellness checks are performed, which are very simple statistical tests, with~the goal of detecting critical failures in the entropy source.
  \item Cryptographic processing of the randomness with AES is performed.
\end{enumerate}

A user requests randomness from the true-RNG via the RDSEED instruction. Using Intel's true-RNG, we cannot directly output raw randomness from the entropy source; the best available option is RDSEED, which includes some level of post-processing (as described above). Independent analyses of the Intel true-RNG's quality, including~\cite{hamburg2012analysis,shrimpton2015provable}, report a min-entropy rate for RDSEED of around 0.65, closely matching our result (see \Cref{sec:rng-analysis}).

\subsection{IDQ Quantis~QRNG}
\label{app:quantis-qrng}
The IDQ Quantis (USB) is a QRNG based on photons hitting a 50:50 beam splitter and being detected in position 0 (reflected) and 1 (transmitted)\@.
In principle, if all components are accurately modelled and the device is shielded from any outside influence, the~output is perfectly random numbers due to the laws of quantum mechanics.  
We refer the reader to the IDQ Quantis QRNG brochure for further details of its construction~\cite{quantique2004quantis}\@.

\section{Initial RNG~Analysis}
\label{app:rng-benchmarking}
\unskip

\subsection{Full Results: Statistical~Testing}
\label{app:initial-rng-testing}
\vspace{-6pt}
\begin{table}[H]
  \caption{{The} 
 number of failed tests for the raw output from the 32-bit LFSR.
  Note that only 127/920 PractRand tests were run due to the numerous failings in the $2^{25}$ byte case. 
  In cells with multiple entries, failed tests are on the left and suspicious tests (when applicable) are on the right in parentheses.}
  \newcolumntype{C}{>{\centering\arraybackslash}X}
\begin{tabularx}{\textwidth}{CcccCCCC}
  \toprule
  \textbf{RNG}     & \begin{tabular}[c]{@{}c@{}}\textbf{NIST}\\ \textbf{(15)}\end{tabular} & \begin{tabular}[c]{@{}c@{}}\textbf{Diehard}\\ \textbf{(18)}\end{tabular} & \begin{tabular}[c]{@{}c@{}}\textbf{ENT}\\ \textbf{(6)}\end{tabular} & \begin{tabular}[c]{@{}c@{}}\textbf{SmallCrush}\\ \textbf{(15)}\end{tabular} & \begin{tabular}[c]{@{}c@{}}\textbf{Alphabit}\\ \textbf{(17)}\end{tabular} & \begin{tabular}[c]{@{}c@{}}\textbf{Rabbit}\\ \textbf{(40)}\end{tabular} & \begin{tabular}[c]{@{}c@{}}\textbf{PractRand}\\ \textbf{(920)}\end{tabular} \\ \midrule
  LFSR 1  & 2                                                    & 8  (0)                                                   & 1                                                 & 10                                                        & 15                                                      & 28                                                    & 23  (5)                                                  \\ \midrule
  LFSR 2  & 2                                                    & 8  (0)                                                   & 1                                                 & 10                                                        & 14                                                      & 27                                                    & 25  (3)                                                  \\ \midrule
  LFSR 3  & 2                                                    & 8  (2)                                                   & 1                                                 & 11                                                        & 15                                                      & 26                                                    & 24  (4)                                                  \\ \midrule
  LFSR 4  & 2                                                    & 8  (0)                                                   & 1                                                 & 10                                                        & 15                                                      & 25                                                    & 23  (5)                                                  \\ \midrule
  LFSR 5  & 2                                                    & 8  (2)                                                   & 1                                                 & 10                                                        & 15                                                      & 25                                                    & 25  (4)                                                  \\ \midrule
  LFSR 6  & 2                                                    & 8  (1)                                                   & 1                                                 & 10                                                        & 15                                                      & 27                                                    & 23  (5)                                                  \\ \midrule
  LFSR 7  & 2                                                    & 8  (0)                                                   & 1                                                 & 10                                                        & 14                                                      & 27                                                    & 24  (6)                                                  \\ \midrule
  LFSR 8  & 2                                                    & 8  (0)                                                   & 1                                                 & 10                                                        & 15                                                      & 25                                                    & 25  (3)                                                  \\ \midrule
  LFSR 9  & 2                                                    & 8  (0)                                                   & 1                                                 & 10                                                        & 14                                                      & 26                                                    & 21  (6)                                                 \\ \midrule
  LFSR 10 & 2                                                    & 8  (1)                                                   & 1                                                 & 10                                                        & 14                                                      & 26                                                    & 23  (4)                                                  \\ \midrule
  \textbf{{Total} 
}     & \textbf{20}                                           & \textbf{80  (6)}                                          & \textbf{10}                                        & \textbf{101}                                                & \textbf{146}                                              & \textbf{262}                                            & \textbf{236  (45)}                                          \\ \bottomrule
  \end{tabularx}
  
\end{table}
\unskip

\begin{table}[H]
\tablesize{\footnotesize}
\caption{{The} 
 number of failed tests for the raw output from RDSEED. 
  In cells with multiple entries, failed tests are on the left and suspicious tests (when applicable) are on the right in parentheses.}
   \newcolumntype{C}{>{\centering\arraybackslash}X}
\begin{tabularx}{\textwidth}{CcccCCcC}
    \toprule
    \textbf{RNG}     & \begin{tabular}[c]{@{}c@{}}\textbf{NIST}\\ \textbf{(15)}\end{tabular} & \begin{tabular}[c]{@{}c@{}}\textbf{Diehard}\\ \textbf{(18)}\end{tabular} & \begin{tabular}[c]{@{}c@{}}\textbf{ENT}\\ \textbf{(6)}\end{tabular} & \begin{tabular}[c]{@{}c@{}}\textbf{SmallCrush}\\ \textbf{(15)}\end{tabular} & \begin{tabular}[c]{@{}c@{}}\textbf{Alphabit}\\ \textbf{(17)}\end{tabular} & \begin{tabular}[c]{@{}c@{}}\textbf{Rabbit}\\ \textbf{(40)}\end{tabular} & \begin{tabular}[c]{@{}c@{}}\textbf{PractRand}\\ \textbf{(920)}\end{tabular} \\ \midrule
    RDSEED 1                  &  0                                & 0  (0)                                                                        & 0                                                                      & 1                                                                              & 0                                                                             & 0                                                                          & 0  (1)                                                                        \\ \midrule
    RDSEED 2                  &  0                                & 0  (1)                                                                        & 0                                                                      & 0                                                                              & 0                                                                             & 0                                                                          & 0  (8)                                                                        \\ \midrule
    RDSEED 3                  &  0                                & 0  (0)                                                                        & 0                                                                      & 0                                                                              & 0                                                                             & 0                                                                          & 0  (0)                                                                         \\ \midrule
    RDSEED 4                  &  0                                & 0  (1)                                                                        & 0                                                                      & 0                                                                              & 0                                                                             & 0                                                                          & 0  (1)                                                                         \\ \midrule
    RDSEED 5                  &  0                                & 0  (1)                                                                        & 0                                                                      & 0                                                                              & 0                                                                             & 1                                                                          & 0  (1)                                                                         \\ \midrule
    RDSEED 6                  &  0                                & 0  (0)                                                                        & 0                                                                      & 0                                                                              & 0                                                                             & 1                                                                          & 0  (0)                                                                         \\ \midrule
    RDSEED 7                  &  0                                & 0  (1)                                                                        & 0                                                                      & 0                                                                              & 0                                                                             & 0                                                                          & 0  (0)                                                                         \\ \midrule
    RDSEED 8                  &  0                                & 0  (1)                                                                        & 0                                                                      & 0                                                                              & 0                                                                             & 0                                                                          & 0  (0)                                                                       \\ \midrule
    RDSEED 9                  &  0                                & 0  (0)                                                                        & 0                                                                      & 0                                                                              & 0                                                                             & 0                                                                          & 0  (1)                                                                         \\ \midrule
    RDSEED 10                 &  0                                & 0  (2)                                                                        & 0                                                                      & 0                                                                              & 0                                                                             & 0                                                                          & 0  (0)                                                                         \\ \midrule
    \textbf{Total}        & \textbf{0}                                           & \textbf{0  (7)}                                          & \textbf{0}                                        & \textbf{1}                                                & \textbf{0}                                              & \textbf{2}                                            & \textbf{0  (12)}                                           \\ \bottomrule
  \end{tabularx}
\end{table}
\unskip

\begin{table}[H]
\tablesize{\footnotesize}
  \caption{{The} 
 number of failed tests for the raw output from IDQ Quantis. 
  In cells with multiple entries, failed tests are on the left and suspicious tests (when applicable) are on the right in parentheses.}
		\newcolumntype{C}{>{\centering\arraybackslash}X}
		\begin{tabularx}{\textwidth}{Cccccccc}
  \toprule
  \textbf{RNG}     & \begin{tabular}[c]{@{}c@{}}\textbf{NIST}\\ \textbf{(15)}\end{tabular} & \begin{tabular}[c]{@{}c@{}}\textbf{Diehard}\\ \textbf{(18)}\end{tabular} & \begin{tabular}[c]{@{}c@{}}\textbf{ENT}\\ \textbf{(6)}\end{tabular} & \begin{tabular}[c]{@{}c@{}}\textbf{SmallCrush}\\ \textbf{(15)}\end{tabular} & \begin{tabular}[c]{@{}c@{}}\textbf{Alphabit}\\ \textbf{(17)}\end{tabular} & \begin{tabular}[c]{@{}c@{}}\textbf{Rabbit}\\ \textbf{(40)}\end{tabular} & \begin{tabular}[c]{@{}c@{}}\textbf{PractRand}\\ \textbf{(920)}\end{tabular} \\ \midrule
  IDQ Quantis 1             & 0                                 & 0  (0)                                                                        & 1                                                                      & 0                                                                              & 3                                                                             & 5                                                                          & 0 (3)                                                                         \\ \midrule
  IDQ Quantis 2             & 0                                 & 0  (1)                                                                        & 1                                                                      & 0                                                                              & 4                                                                             & 5                                                                          & 0  (2)                                                                         \\ \midrule
  IDQ Quantis 3             & 0                                 & 0  (0)                                                                        & 1                                                                      & 0                                                                              & 2                                                                             & 4                                                                          & 0  (11)                                                                         \\ \midrule
  IDQ Quantis 4             & 0                                & 0  (1)                                                                        & 1                                                                      & 0                                                                              & 3                                                                             & 5                                                                          & 0  (0)                                                                         \\
 \bottomrule
  \end{tabularx}
\end{table}

\begin{table}[H]\ContinuedFloat
\tablesize{\footnotesize}
  \caption{\emph{Cont.}}
		\newcolumntype{C}{>{\centering\arraybackslash}X}
		\begin{tabularx}{\textwidth}{Cccccccc}
  \toprule
  \textbf{RNG}     & \begin{tabular}[c]{@{}c@{}}\textbf{NIST}\\ \textbf{(15)}\end{tabular} & \begin{tabular}[c]{@{}c@{}}\textbf{Diehard}\\ \textbf{(18)}\end{tabular} & \begin{tabular}[c]{@{}c@{}}\textbf{ENT}\\ \textbf{(6)}\end{tabular} & \begin{tabular}[c]{@{}c@{}}\textbf{SmallCrush}\\ \textbf{(15)}\end{tabular} & \begin{tabular}[c]{@{}c@{}}\textbf{Alphabit}\\ \textbf{(17)}\end{tabular} & \begin{tabular}[c]{@{}c@{}}\textbf{Rabbit}\\ \textbf{(40)}\end{tabular} & \begin{tabular}[c]{@{}c@{}}\textbf{PractRand}\\ \textbf{(920)}\end{tabular} \\ \midrule
  IDQ Quantis 5             & 0                                 & 0  (1)                                                                        & 1                                                                      & 0                                                                              & 3                                                                             & 5                                                                          & 0  (2)                                                                         \\ \midrule
  IDQ Quantis 6             & 0                                 & 0  (0)                                                                        & 1                                                                      & 0                                                                              & 2                                                                             & 5                                                                          & 0  (1)                                                                         \\ \midrule
  IDQ Quantis 7             & 0                                  & 0  (0)                                                                        & 1                                                                      & 0                                                                              & 5                                                                             & 5                                                                          & 0  (4)                                                                         \\ \midrule
  IDQ Quantis 8             & 0                                 & 0  (1)                                                                        & 1                                                                      & 0                                                                              & 3                                                                             & 6                                                                          & 2  (1)                                                                         \\ \midrule
  IDQ Quantis 9             & 0                                 & 0  (1)                                                                        & 1                                                                      & 0                                                                              & 3                                                                             & 4                                                                          & 0  (4)                                                                         \\ \midrule
  IDQ Quantis 10            & 0                                 & 0  (1)                                                                        & 1                                                                      & 0                                                                              & 6                                                                             & 5                                                                          & 3  (2)                                                                         \\ \midrule
  \textbf{Total}     & \textbf{0}                                           & \textbf{0  (6)}                                          & \textbf{10}                                        & \textbf{0}                                                & \textbf{34}                                              & \textbf{49}                                            & \textbf{5  (28)}                                          \\
 \bottomrule
  \end{tabularx}
\end{table}

\subsection{Full Results: Min-Entropy~Estimators}
\label{app:min-ent-est}
\vspace{-6pt}
\begin{table}[H]
\tablesize{\footnotesize}
  \caption{{Observed} 
 NIST min-entropy estimators for 32-bit LFSR raw~output.}
  \newcolumntype{C}{>{\centering\arraybackslash}X}
\begin{tabularx}{\textwidth}{cCC}
  \toprule
  {\textbf{RNG}}     & \begin{tabular}[c]{@{}c@{}}\textbf{NIST} \textbf{Min-Entropy}  \textbf{Estimator} \textbf{(/byte)}\end{tabular} & \begin{tabular}[c]{@{}c@{}}\textbf{NIST}  \textbf{Min-Entropy} \textbf{Estimator} \textbf{(/bit)}\end{tabular} \\ \midrule
  LFSR 1  & 6.956997                                                                          & 0.869624625                                                                      \\ \midrule
  LFSR 2  & 5.792304                                                                          & 0.724038                                                                         \\ \midrule
  LFSR 3  & 7.161811                                                                          & 0.895226375                                                                      \\ \midrule
  LFSR 4  & 6.638405                                                                          & 0.829800625                                                                      \\ \midrule
  LFSR 5  & 7.353758                                                                          & 0.91921975                                                                       \\ \midrule
  LFSR 6  & 7.121091                                                                          & 0.890136375                                                                      \\ \midrule
  LFSR 7  & 7.213483                                                                          & 0.901685375                                                                      \\ \midrule
  LFSR 8  & 7.188889                                                                          & 0.898611125                                                                      \\ \midrule
  LFSR 9  & 6.638383                                                                          & 0.829797875                                                                      \\ \midrule
  LFSR 10 & 6.638383                                                                          & 0.829797875                                                                      \\ \midrule
  \textbf{Average}        & \textbf{6.8703504}                                           & \textbf{0.8587938}                                                                                        \\ \bottomrule
  \end{tabularx}
  \label{table:min-ent-lfsr}
\end{table}
\unskip

\begin{table}[H]
\tablesize{\footnotesize}
  \caption{{Observed} 
 NIST min-entropy estimators for RDSEED raw~output.}
 \newcolumntype{C}{>{\centering\arraybackslash}X}
\begin{tabularx}{\textwidth}{cCC}
  \toprule
 {\textbf{RNG}}     & \begin{tabular}[c]{@{}c@{}}\textbf{NIST} \textbf{Min-Entropy}  \textbf{Estimator} \textbf{(/byte)}\end{tabular} & \begin{tabular}[c]{@{}c@{}}\textbf{NIST}  \textbf{Min-Entropy} \textbf{Estimator} \textbf{(/bit)}\end{tabular} \\ \midrule
  RDSEED 1  & 6.737815                                                                           & 0.842226875                                                                      \\ \midrule
  RDSEED 2  & 6.530758                                                                           & 0.81634475                                                                       \\ \midrule
  RDSEED 3  & 6.846048                                                                           & 0.855756                                                                         \\ \midrule
  RDSEED 4  & 6.995008                                                                           & 0.874376                                                                         \\ \midrule
  RDSEED 5  & 6.861225                                                                           & 0.857653125                                                                      \\ \midrule
  RDSEED 6  & 7.086914                                                                           & 0.88586425                                                                       \\ \midrule
  RDSEED 7  & 6.638399                                                                           & 0.829799875                                                                      \\ \midrule
  RDSEED 8  & 7.024343                                                                           & 0.878042875                                                                      \\ \midrule
  RDSEED 9  & 6.747707                                                                           & 0.843463375                                                                      \\ \midrule
  RDSEED 10 & 6.724567                                                                           & 0.840570875                                                                      \\ \midrule
  \textbf{Average}        & \textbf{6.8192784}                                           & \textbf{0.8524098}                                                                                        \\ \bottomrule
  \end{tabularx}
  \label{table:min-ent-RDSEED}
\end{table}
\unskip

\begin{table}[H]
\tablesize{\footnotesize}
    \caption{{Observed} 
 NIST min-entropy estimators for IDQ Quantis raw~output.}
    \newcolumntype{C}{>{\centering\arraybackslash}X}
\begin{tabularx}{\textwidth}{cCC}
  \toprule
{\textbf{RNG}}     & \begin{tabular}[c]{@{}c@{}}\textbf{NIST} \textbf{Min-Entropy}  \textbf{Estimator} \textbf{(/byte)}\end{tabular} & \begin{tabular}[c]{@{}c@{}}\textbf{NIST}  \textbf{Min-Entropy} \textbf{Estimator} \textbf{(/bit)}\end{tabular} \\ \midrule
    IDQ Quantis 1  & 7.149988                                                                           & 0.8937485                                                                        \\ \midrule
    IDQ Quantis 2  & 7.142161                                                                           & 0.892770125                                                                      \\ \midrule
    IDQ Quantis 3  & 7.152185                                                                           & 0.894023125                                                                      \\ \midrule
    IDQ Quantis 4  & 7.088475                                                                           & 0.886059375                                                                      
 \\ \bottomrule
    \end{tabularx}
    \label{table:min-ent-idq}
\end{table}

\begin{table}[H]\ContinuedFloat
\tablesize{\footnotesize}
    \caption{\emph{Cont.}}
    \newcolumntype{C}{>{\centering\arraybackslash}X}
\begin{tabularx}{\textwidth}{cCC}
  \toprule
{\textbf{RNG}}     & \begin{tabular}[c]{@{}c@{}}\textbf{NIST} \textbf{Min-Entropy}  \textbf{Estimator} \textbf{(/byte)}\end{tabular} & \begin{tabular}[c]{@{}c@{}}\textbf{NIST}  \textbf{Min-Entropy} \textbf{Estimator} \textbf{(/bit)}\end{tabular} \\ \midrule
    IDQ Quantis 5  & 7.161971                                                                           & 0.895246375                                                                      \\ \midrule
    IDQ Quantis 6  & 7.169887                                                                           & 0.896235875                                                                      \\ \midrule
    IDQ Quantis 7  & 7.260033                                                                           & 0.907504125                                                                      \\ \midrule
    IDQ Quantis 8  & 7.188102                                                                           & 0.89851275                                                                       \\ \midrule
    IDQ Quantis 9  & 7.115609                                                                           & 0.889451125                                                                      \\ \midrule
    IDQ Quantis 10 & 7.142009                                                                           & 0.892751125                                                                      \\ \midrule
    \textbf{Average}        & \textbf{7.157042}                                           & \textbf{0.89463025}                                                                                        \\ \bottomrule
    \end{tabularx}
    \label{table:min-ent-idq}
\end{table}

\subsubsection{Deriving a Min-Entropy Lower~Bound}
\label{app:min-ent-bound}
In this subsection, we derive a min-entropy lower bound from the observed NIST min-entropy estimators in Table~\ref{table:min-ent-est}\@. 
We use subscripts to index a single min-entropy estimator test and superscript to index the RNG which the variable refers to. 
Let $\mathsf{est}_i$ denote the $i$th observed NIST min-entropy estimate per bit for a test, $i = 1, \ldots, 10$, and~$\overline{\mathsf{est}}$ the average estimate per bit.
The sample standard deviation $\sigma$ (using Bessel's correction) is given by
\begin{align} \label{eq:sigma}
\sigma^{\mathsf{RNG}} = \sqrt{\frac{1}{n-1}\sum_{i=1}^n (\overline{\mathsf{est}}^{\mathsf{RNG}} - \mathsf{est_i^{RNG}})^2}\ .
\end{align} 
We compute the lower bound for the min-entropy rate, $\alpha^{\mathsf{RNG}}$, including a finite statistics correction term to lower-bound the true estimated min-entropy rate of each RNG with a high probability. 
Specifically, we desire
\begin{align} \label{eps_est} \Pr(\mathsf{est_i}^{\mathsf{RNG}} < \alpha^{\mathsf{RNG}}) = \epsilon_{\mathsf{est}} < 2^{-32}, \end{align}  
where $\mathsf{est_i}$ is the $i$th NIST min-entropy estimator for a specific RNG. 
Selecting
\begin{align} \alpha^{\mathsf{RNG}} = {\overline{\mathsf{est}}^{\mathsf{RNG}}} - 7 \sigma^{\mathsf{RNG}} \end{align}
where $\mathrm{\overline{est}^{\mathsf{RNG}}}$ is the average NIST min-entropy estimator for the RNG and $\sigma^{\mathsf{RNG}}$ is the observed sample standard deviation that satisfies \Cref{eps_est}, giving $\epsilon_{\mathsf{est}} \approx 2^{-39}$\@. Here, we have made the assumption that the NIST min-entropy estimator results are normally distributed (which we believe is reasonable due to each test sample being generated a significant time apart) and used the standard probability density function for normally distributed~variables. 

\section{Deterministic Extraction in~Detail}
\label{app:vn-debiased}

\subsection{Full~Results}
\label{app:det-extract}
\vspace{-6pt}
\begin{table}[H]
\tablesize{\footnotesize}
  \caption{{The} 
 number of failed tests for the output of the 32-bit LFSR after post-processing with the \VonNeumann extractor.
  Note that only 127/920 PractRand tests were run due to the numerous failings in the $2^{25}$ byte case. 
  In cells with multiple entries, failed tests are on the left and suspicious tests (when applicable) are on the right in parentheses.}
  \newcolumntype{C}{>{\centering\arraybackslash}X}
		\begin{tabularx}{\textwidth}{cCCCCCCC}
  \toprule
  \textbf{RNG}              & \begin{tabular}[c]{@{}c@{}}\textbf{NIST}\\ \textbf{(15)}\end{tabular} & \begin{tabular}[c]{@{}c@{}}\textbf{Diehard}\\ \textbf{(18)}\end{tabular} & \begin{tabular}[c]{@{}c@{}}\textbf{ENT}\\ \textbf{(6)}\end{tabular} & \begin{tabular}[c]{@{}c@{}}\textbf{SmallCrush}\\ \textbf{(15)}\end{tabular} & \begin{tabular}[c]{@{}c@{}}\textbf{Alphabit}\\ \textbf{(17)}\end{tabular} & \begin{tabular}[c]{@{}c@{}}\textbf{Rabbit}\\ \textbf{(40)}\end{tabular} & \begin{tabular}[c]{@{}c@{}}\textbf{PractRand}\\ \textbf{(920)}\end{tabular} \\ \midrule
  LFSR VN 1 & 5                                                    & 2  (2)                                                   & 1                                                 & 4                                                         & 16                                                      & 21                                                    & 19  (12)                                                 \\ \midrule
  LFSR VN 2 & 5                                                    & 2  (0)                                                   & 1                                                 & 3                                                         & 15                                                      & 21                                                    & 21  (11)                                                  \\ \midrule
  LFSR VN 3 & 5                                                    & 2  (1)                                                   & 1                                                 & 3                                                         & 15                                                      & 22                                                    & 22  (10)                                                  \\
\bottomrule
  \end{tabularx}
\end{table}

\begin{table}[H]\ContinuedFloat
\tablesize{\footnotesize}
  \caption{\emph{Cont.}}
  \newcolumntype{C}{>{\centering\arraybackslash}X}
		\begin{tabularx}{\textwidth}{cCCCCCCC}
  \toprule
  \textbf{RNG}              & \begin{tabular}[c]{@{}c@{}}\textbf{NIST}\\ \textbf{(15)}\end{tabular} & \begin{tabular}[c]{@{}c@{}}\textbf{Diehard}\\ \textbf{(18)}\end{tabular} & \begin{tabular}[c]{@{}c@{}}\textbf{ENT}\\ \textbf{(6)}\end{tabular} & \begin{tabular}[c]{@{}c@{}}\textbf{SmallCrush}\\ \textbf{(15)}\end{tabular} & \begin{tabular}[c]{@{}c@{}}\textbf{Alphabit}\\ \textbf{(17)}\end{tabular} & \begin{tabular}[c]{@{}c@{}}\textbf{Rabbit}\\ \textbf{(40)}\end{tabular} & \begin{tabular}[c]{@{}c@{}}\textbf{PractRand}\\ \textbf{(920)}\end{tabular} \\ \midrule
  LFSR VN 4 & 5                                                    & 2  (2)                                                   & 1                                                 & 3                                                         & 15                                                      & 22                                                    & 19  (12)                                                \\ \midrule
  LFSR VN 5 & 5                                                    & 2  (0)                                                   & 1                                                 & 5                                                         & 15                                                      & 20                                                    & 19  (12)                                                  \\ \midrule
  \textbf{Total}     & \textbf{25}                                           & \textbf{10  (5)}                                          & \textbf{5}                                        & \textbf{18}                                                & \textbf{76}                                              & \textbf{106}                                            & \textbf{100  (57)}                                          \\ 
\bottomrule
  \end{tabularx}
\end{table}
\unskip

\begin{table}[H]
\tablesize{\scriptsize}
  \caption{{The} 
 number of failed tests for the output of RDSEED after post-processing with the \VonNeumann extractor.
  In cells with multiple entries, failed tests are on the left and suspicious tests (when applicable) are on the right in parentheses.}
  \newcolumntype{C}{>{\centering\arraybackslash}X}
		\begin{tabularx}{\textwidth}{cCCCCCCC}
  \toprule
  \textbf{RNG}              & \begin{tabular}[c]{@{}c@{}}\textbf{NIST}\\ \textbf{(15)}\end{tabular} & \begin{tabular}[c]{@{}c@{}}\textbf{Diehard}\\ \textbf{(18)}\end{tabular} & \begin{tabular}[c]{@{}c@{}}\textbf{ENT}\\ \textbf{(6)}\end{tabular} & \begin{tabular}[c]{@{}c@{}}\textbf{SmallCrush}\\ \textbf{(15)}\end{tabular} & \begin{tabular}[c]{@{}c@{}}\textbf{Alphabit}\\ \textbf{(17)}\end{tabular} & \begin{tabular}[c]{@{}c@{}}\textbf{Rabbit}\\ \textbf{(40)}\end{tabular} & \begin{tabular}[c]{@{}c@{}}\textbf{PractRand}\\ \textbf{(920)}\end{tabular} \\ \midrule
  RDSEED VN 1 & 0                                                    & 0  (0)                                                   & 0                                                 & 0                                                         & 0                                                       & 0                                                     & 0  (0)                                                    \\ \midrule
  RDSEED VN 2 & 0                                                    & 0  (0)                                                   & 0                                                 & 0                                                         & 0                                                       & 0                                                     & 0  (0)                                                   \\ \midrule
  RDSEED VN 3 & 0                                                    & 0  (0)                                                   & 0                                                 & 0                                                         & 0                                                       & 0                                                     & 0  (1)                                                   \\ \midrule
  RDSEED VN 4 & 0                                                    & 0  (1)                                                   & 0                                                 & 0                                                         & 0                                                       & 0                                                     & 0  (1)                                                   \\ \midrule
  RDSEED VN 5 & 0                                                    & 0  (1)                                                   & 0                                                 & 0                                                         & 0                                                       & 1                                                     & 0  (0)                                                   \\ \midrule
  \textbf{Total}     & \textbf{0}                                           & \textbf{0  (2)}                                          & \textbf{0}                                        & \textbf{0}                                                & \textbf{0}                                              & \textbf{1}                                            & \textbf{0  (2)}                                          \\ \bottomrule
  \end{tabularx}
\end{table}
\unskip

\begin{table}[H]
  \tablesize{\footnotesize}
  \caption{{The} 
 number of failed tests for the output of IDQ Quantis after post-processing with the \VonNeumann extractor.
  In cells with multiple entries, failed tests are on the left and suspicious tests (when applicable) are on the right in parentheses.}
		\newcolumntype{C}{>{\centering\arraybackslash}X}
		\begin{tabularx}{\textwidth}{Cccccccc}
  \toprule
  \textbf{RNG}              & \begin{tabular}[c]{@{}c@{}}\textbf{NIST}\\ \textbf{(15)}\end{tabular} & \begin{tabular}[c]{@{}c@{}}\textbf{Diehard}\\ \textbf{(18)}\end{tabular} & \begin{tabular}[c]{@{}c@{}}\textbf{ENT}\\ \textbf{(6)}\end{tabular} & \begin{tabular}[c]{@{}c@{}}\textbf{SmallCrush}\\ \textbf{(15)}\end{tabular} & \begin{tabular}[c]{@{}c@{}}\textbf{Alphabit}\\ \textbf{(17)}\end{tabular} & \begin{tabular}[c]{@{}c@{}}\textbf{Rabbit}\\ \textbf{(40)}\end{tabular} & \begin{tabular}[c]{@{}c@{}}\textbf{PractRand}\\ \textbf{(920)}\end{tabular} \\ \midrule
  IDQ Quantis VN 1 &  0                                                   & 0  (0)                                                   & 0                                                 & 0                                                         & 0                                                       & 0                                                     & 0  (1)                                                   \\ \midrule
  IDQ Quantis VN 2 &  0                                                   & 0  (0)                                                   & 0                                                 & 0                                                         & 0                                                       & 2                                                     & 0  (0)                                                    \\ \midrule
  IDQ Quantis VN 3 &  0                                                   & 0  (0)                                                   & 0                                                 & 0                                                         & 0                                                       & 0                                                     & 0  (1)                                                   \\ \midrule
  IDQ Quantis VN 4 &  2                                                   & 0  (1)                                                   & 0                                                 & 0                                                         & 0                                                       & 1                                                     & 0  (0)                                                    \\ \midrule
  IDQ Quantis VN 5 &  2                                                   & 0  (0)                                                   & 0                                                 & 0                                                         & 0                                                       & 0                                                     & 0  (1)                                                   \\ \midrule
  \textbf{Total}     & \textbf{4}                                           & \textbf{0  (1)}                                          & \textbf{0}                                        & \textbf{0}                                                & \textbf{0}                                              & \textbf{3}                                            & \textbf{0  (3)}                                          \\ \bottomrule
  \end{tabularx}
\end{table}

\section{Seeded Extraction in~Detail}
\label{app:nist-se}

\subsection{Full~Results}
\label{app:nist-se-beacon}
\vspace{-6pt}
\begin{table}[H]
  \caption{{Statistical} 
 test results for the 32-bit LFSR as the weak input source to the strong seeded randomness \Circulant extractor. 
  The seed is randomness generated from the NIST Randomness Beacon.
  In cells with multiple entries, failed tests are on the left and suspicious tests (when applicable) are on the right in parentheses.}
 \newcolumntype{C}{>{\centering\arraybackslash}X}
		\begin{tabularx}{\textwidth}{Cccccccc}
  \toprule
  \textbf{RNG}              & \begin{tabular}[c]{@{}c@{}}\textbf{NIST}\\ \textbf{(15)}\end{tabular} & \begin{tabular}[c]{@{}c@{}}\textbf{Diehard}\\ \textbf{(18)}\end{tabular} & \begin{tabular}[c]{@{}c@{}}\textbf{ENT}\\ \textbf{(6)}\end{tabular} & \begin{tabular}[c]{@{}c@{}}\textbf{SmallCrush}\\ \textbf{(15)}\end{tabular} & \begin{tabular}[c]{@{}c@{}}\textbf{Alphabit}\\ \textbf{(17)}\end{tabular} & \begin{tabular}[c]{@{}c@{}}\textbf{Rabbit}\\ \textbf{(40)}\end{tabular} & \begin{tabular}[c]{@{}c@{}}\textbf{PractRand}\\ \textbf{(920)}\end{tabular} \\ \midrule
  LFSR NIST SE 1 &  0                                                   & 0  (2)                                                   & 0                                                 & 0                                                         & 0                                                       & 0                                                     & 0    (2)                                                   \\ \midrule
  LFSR NIST SE 2 &  0                                                   & 0  (0)                                                   & 0                                                 & 0                                                         & 0                                                       & 0                                                     & 0    (1)                                                   \\ \midrule
  LFSR NIST SE 3 &  0                                                   & 0  (0)                                                   & 0                                                 & 0                                                         & 0                                                       & 0                                                     & 0    (1)                                                   \\ \midrule
  LFSR NIST SE 4 &  0                                                   & 0  (1)                                                   & 0                                                 & 0                                                         & 0                                                       & 0                                                     & 0   (0)                                                   \\ \midrule
  LFSR NIST SE 5 &  0                                                   & 0  (0)                                                   & 0                                                 & 0                                                         & 0                                                       & 0                                                     & 0    (2)                                                   \\ \midrule
  \textbf{Total}   & \textbf{0}                                           & \textbf{0  (3)}                                          & \textbf{0}                                        & \textbf{0}                                                & \textbf{0}                                              & \textbf{0}                                            & \textbf{0    (6)}                                          \\ \bottomrule
  \end{tabularx}
\end{table}
\unskip

\begin{table}[H]
 \caption{{Statistical} 
 test results for RDSEED as the weak input source to the strong seeded randomness \Circulant extractor. 
  The seed is randomness generated from the NIST Randomness Beacon.
  In cells with multiple entries, failed tests are on the left and suspicious tests (when applicable) are on the right in parentheses.}
 \newcolumntype{C}{>{\centering\arraybackslash}X}
		\begin{tabularx}{\textwidth}{Cccccccc}
  \toprule
  \textbf{RNG}              & \begin{tabular}[c]{@{}c@{}}\textbf{NIST}\\ \textbf{(15)}\end{tabular} & \begin{tabular}[c]{@{}c@{}}\textbf{Diehard}\\ \textbf{(18)}\end{tabular} & \begin{tabular}[c]{@{}c@{}}\textbf{ENT}\\ \textbf{(6)}\end{tabular} & \begin{tabular}[c]{@{}c@{}}\textbf{SmallCrush}\\ \textbf{(15)}\end{tabular} & \begin{tabular}[c]{@{}c@{}}\textbf{Alphabit}\\ \textbf{(17)}\end{tabular} & \begin{tabular}[c]{@{}c@{}}\textbf{Rabbit}\\ \textbf{(40)}\end{tabular} & \begin{tabular}[c]{@{}c@{}}\textbf{PractRand}\\ \textbf{(920)}\end{tabular} \\ \midrule
  RDSEED NIST SE 1 & 0                                                    & 0  (3)                                                   & 0                                                 & 0                                                         & 0                                                       & 0                                                     & 0    (4)                                                   \\ \midrule
  RDSEED NIST SE 2 & 0                                                    & 0  (1)                                                   & 0                                                 & 0                                                         & 0                                                       & 0                                                     & 0    (2)                                                   \\ \midrule
  RDSEED NIST SE 3 & 0                                                    & 0  (1)                                                   & 0                                                 & 0                                                         & 0                                                       & 0                                                     & 0    (0)                                                   \\ \midrule
  RDSEED NIST SE 4 & 0                                                    & 0  (0)                                                   & 0                                                 & 0                                                         & 0                                                       & 0                                                     & 0    (1)                                                   \\ \midrule
  RDSEED NIST SE 5 & 0                                                    & 0  (2)                                                   & 0                                                 & 0                                                         & 0                                                       & 0                                                     & 0  (0)                                                     \\ \midrule
  \textbf{Total}     & \textbf{0}                                           & \textbf{0  (7)}                                          & \textbf{0}                                        & \textbf{0}                                                & \textbf{0}                                              & \textbf{0}                                            & \textbf{0   (7)}                                          \\ \bottomrule
  \end{tabularx}
 
\end{table}
\unskip

\begin{table}[H]
  \caption{{Statistical} 
 test results for IDQ Quantis as the weak input source to the strong seeded randomness \Circulant extractor. 
  The seed is randomness generated from the NIST Randomness Beacon.
  In cells with multiple entries, failed tests are on the left and suspicious tests (when applicable) are on the right in parentheses.}
  \newcolumntype{C}{>{\centering\arraybackslash}X}
		\begin{tabularx}{\textwidth}{Cccccccc}
  \toprule
  \textbf{RNG}              & \begin{tabular}[c]{@{}c@{}}\textbf{NIST}\\ \textbf{(15)}\end{tabular} & \begin{tabular}[c]{@{}c@{}}\textbf{Diehard}\\ \textbf{(18)}\end{tabular} & \begin{tabular}[c]{@{}c@{}}\textbf{ENT}\\ \textbf{(6)}\end{tabular} & \begin{tabular}[c]{@{}c@{}}\textbf{SmallCrush}\\ \textbf{(15)}\end{tabular} & \begin{tabular}[c]{@{}c@{}}\textbf{Alphabit}\\ \textbf{(17)}\end{tabular} & \begin{tabular}[c]{@{}c@{}}\textbf{Rabbit}\\ \textbf{(40)}\end{tabular} & \begin{tabular}[c]{@{}c@{}}\textbf{PractRand}\\ \textbf{(920)}\end{tabular} \\ \midrule
  IDQ Quantis NIST SE 1 & 0                                                    & 0  (0)                                                   & 0                                                 & 0                                                         & 0                                                       & 0                                                     & 0    (3)                                                   \\ \midrule
  IDQ Quantis NIST SE 2 & 0                                                    & 0  (0)                                                   & 0                                                 & 0                                                         & 0                                                       & 1                                                     & 0  (0)                                                     \\ \midrule
  IDQ Quantis NIST SE 3 & 0                                                    & 0  (2)                                                   & 0                                                 & 0                                                         & 0                                                       & 1                                                     & 0  (0)                                                     \\ \midrule
  IDQ Quantis NIST SE 4 & 0                                                    & 0  (0)                                                   & 0                                                 & 0                                                         & 0                                                       & 0                                                     & 0   (2)                                                   \\ \midrule
  IDQ Quantis NIST SE 5 & 0                                                    & 0  (0)                                                   & 0                                                 & 0                                                         & 0                                                       & 0                                                     & 0  (0)                                                   \\ \midrule
  \textbf{Total}          & \textbf{0}                                           & \textbf{0  (2)}                                          & \textbf{0}                                        & \textbf{0}                                                & \textbf{0}                                              & \textbf{2}                                            & \textbf{0  (5)}                                          \\ \bottomrule
  \end{tabularx}
\end{table}

\section{Two-Source Extraction in~Detail}
\label{app:2-source}
\unskip

\subsection{Two-Source Extraction with a Single~RNG}
\label{app:2-source-independence}
In this subsection, we test the use of two strings from each RNG as the inputs to a strong two-source extractor.
For near-perfect randomness to be generated, the~unique strings from the RNG must be independent; otherwise, this violates some of the assumptions of this level.
Due to the limitations of the \Circulant strong two-source extractor, we are unable to perform this step for the 
LFSR, since the min-entropy lower bound derived in Table~\ref{table:min-ent-est} is too~low. 

\begin{table}[H]
 \tablesize{\scriptsize}
  \caption{{This} 
 table gives the sum of statistical tests failed for $5 \times 10$Gbit samples from each RNG, after~strong 2-source extraction taking strings of randomness from the same RNG and assuming independence.
  Note: The 32-bit LFSR does not generate any output in this setting, since its min-entropy is too low.
  In cells with multiple entries, failed tests are on the left and suspicious tests (when applicable) are on the right in parentheses.}
  \newcolumntype{C}{>{\centering\arraybackslash}X}
\begin{tabularx}{\textwidth}{cCCCCCCC}
  \toprule
  \textbf{RNG}         & \begin{tabular}[c]{@{}c@{}}\textbf{NIST}\\ \textbf{(75)}\end{tabular} & \begin{tabular}[c]{@{}c@{}}\textbf{Diehard}\\ \textbf{(90)}\end{tabular} & \begin{tabular}[c]{@{}c@{}}\textbf{ENT}\\ \textbf{(30)}\end{tabular} & \begin{tabular}[c]{@{}c@{}}\textbf{SmallCrush}\\ \textbf{(75)}\end{tabular} & \begin{tabular}[c]{@{}c@{}}\textbf{Alphabit}\\ \textbf{(85)}\end{tabular} & \begin{tabular}[c]{@{}c@{}}\textbf{Rabbit}\\ \textbf{(200)}\end{tabular} & \begin{tabular}[c]{@{}c@{}}\textbf{PractRand}\\ \textbf{(4600)}\end{tabular} \\ \midrule
  32-bit LFSR & -                                                   & -                                                      & -                                                  & -                                                         & -                                                        & -                                                      & -                                                          \\ \midrule
  RDSEED      & 0                                                    & 0  (1)                                                   & 1                                                  & 0                                                         & 0                                                        & 1                                                      & 0  (7)                                                    \\ \midrule
  IDQ Quantis & 0                                                    & 0  (2)                                                   & 0                                                  & 0                                                         & 2                                                        & 1                                                      & 0  (8)                                                    \\ \bottomrule
  \end{tabularx}
\end{table}
\unskip

\begin{table}[H]
 \tablesize{\scriptsize}
  \caption{{Statistical} 
 test results for RDSEED as the weak input source to the strong two-source randomness \Circulant extractor. 
  The seed is randomness generated from RDSEED\@.
  In cells with multiple entries, failed tests are on the left and suspicious tests (when applicable) are on the right \mbox{in parentheses}.}
 \newcolumntype{C}{>{\centering\arraybackslash}X}
		\begin{tabularx}{\textwidth}{Cccccccc}
  \toprule
  \textbf{RNG}              & \begin{tabular}[c]{@{}c@{}}\textbf{NIST}\\ \textbf{(15)}\end{tabular} & \begin{tabular}[c]{@{}c@{}}\textbf{Diehard}\\ \textbf{(18)}\end{tabular} & \begin{tabular}[c]{@{}c@{}}\textbf{ENT}\\ \textbf{(6)}\end{tabular} & \begin{tabular}[c]{@{}c@{}}\textbf{SmallCrush}\\ \textbf{(15)}\end{tabular} & \begin{tabular}[c]{@{}c@{}}\textbf{Alphabit}\\ \textbf{(17)}\end{tabular} & \begin{tabular}[c]{@{}c@{}}\textbf{Rabbit}\\ \textbf{(40)}\end{tabular} & \begin{tabular}[c]{@{}c@{}}\textbf{PractRand}\\ \textbf{(920)}\end{tabular} \\ \midrule
  RDSEED Self 2E 1 &  0                                                   & 0  (1)                                                   & 0                                                 & 0                                                         & 0                                                       & 0                                                     & 0    (2)                                                   \\ \midrule
  RDSEED Self 2E 2 &  0                                                    & 0  (0)                                                   & 0                                                 & 0                                                         & 0                                                       & 1                                                     & 0    (0)                                                   \\ \midrule
  RDSEED Self 2E 3 &  0                                                   & 0  (0)                                                   & 0                                                 & 0                                                         & 0                                                       & 0                                                     & 0    (2)                                                   \\ \midrule
  RDSEED Self 2E 4 &  0                                                   & 0  (0)                                                   & 1                                                 & 0                                                         & 0                                                       & 0                                                     & 0    (3)                                                   \\ \midrule
  RDSEED Self 2E 5 &  0                                                   & 0  (0)                                                   & 0                                                 & 0                                                         & 0                                                       & 0                                                     & 0    (0)                                                   \\ \midrule
  \textbf{Total}   & \textbf{0}                                           & \textbf{0  (1)}                                          & \textbf{1}                                        & \textbf{0}                                                & \textbf{0}                                              & \textbf{1}                                            & \textbf{0  (7)}                                          \\ \bottomrule
  \end{tabularx}
\end{table}
\unskip

\begin{table}[H]
  \caption{{Statistical} 
 test results for IDQ Quantis as the weak input source to the strong two-source randomness \Circulant extractor. 
  The seed is randomness generated from IDQ Quantis.
  In cells with multiple entries, failed tests are on the left and suspicious tests (when applicable) are on the right \mbox{in parentheses}.}
 \newcolumntype{C}{>{\centering\arraybackslash}X}
		\begin{tabularx}{\textwidth}{Cccccccc}
  \toprule
  \textbf{RNG}              & \begin{tabular}[c]{@{}c@{}}\textbf{NIST}\\ \textbf{(15)}\end{tabular} & \begin{tabular}[c]{@{}c@{}}\textbf{Diehard}\\ \textbf{(18)}\end{tabular} & \begin{tabular}[c]{@{}c@{}}\textbf{ENT}\\ \textbf{(6)}\end{tabular} & \begin{tabular}[c]{@{}c@{}}\textbf{SmallCrush}\\ \textbf{(15)}\end{tabular} & \begin{tabular}[c]{@{}c@{}}\textbf{Alphabit}\\ \textbf{(17)}\end{tabular} & \begin{tabular}[c]{@{}c@{}}\textbf{Rabbit}\\ \textbf{(40)}\end{tabular} & \begin{tabular}[c]{@{}c@{}}\textbf{PractRand}\\ \textbf{(920)}\end{tabular} \\ \midrule
  IDQ Quantis Self 2E 1 &  0                                                   & 0  (1)                                                   & 0                                                 & 0                                                         & 0                                                       & 0                                                     & 0    (3)                                                   \\ \midrule
  IDQ Quantis Self 2E 2 &  0                                                   & 0  (0)                                                   & 0                                                 & 0                                                         & 0                                                       & 0                                                     & 0    (2)                                                   \\ \midrule
  IDQ Quantis Self 2E 3 &  0                                                   & 0  (0)                                                   & 0                                                 & 0                                                         & 1                                                       & 1                                                     & 0    (3)                                                   \\ \midrule
  IDQ Quantis Self 2E 4 &  0                                                   & 0  (0)                                                   & 0                                                 & 0                                                         & 1                                                       & 0                                                     & 0    (0)                                                   \\ \midrule
  IDQ Quantis Self 2E 5 &  0                                                   & 0  (1)                                                   & 0                                                 & 0                                                         & 0                                                       & 0                                                     & 0    (0)                                                   \\ \midrule
  \textbf{Total}        & \textbf{0}                                           & \textbf{0  (2)}                                          & \textbf{0}                                        & \textbf{0}                                                & \textbf{2}                                              & \textbf{1}                                            & \textbf{0    (8)}                                          \\ \bottomrule
  \end{tabularx}
\end{table}
\unskip

\subsection{Two-Source Extraction Using the NIST Randomness~Beacon}
\label{app:nist-2e}
\vspace{-6pt}
\begin{table}[H]
 
  \caption{{Statistical} 
 test results for the 32-bit LFSR as the weak input source to the strong two-source randomness \Circulant extractor. 
  The seed is randomness generated from the NIST Randomness Beacon.
  In cells with multiple entries, failed tests are on the left and suspicious tests (when applicable) are on the right in parentheses.}
  \newcolumntype{C}{>{\centering\arraybackslash}X}
		\begin{tabularx}{\textwidth}{Cccccccc}
  \toprule
  \textbf{RNG}              & \begin{tabular}[c]{@{}c@{}}\textbf{NIST}\\ \textbf{(15)}\end{tabular} & \begin{tabular}[c]{@{}c@{}}\textbf{Diehard}\\ \textbf{(18)}\end{tabular} & \begin{tabular}[c]{@{}c@{}}\textbf{ENT}\\ \textbf{(6)}\end{tabular} & \begin{tabular}[c]{@{}c@{}}\textbf{SmallCrush}\\ \textbf{(15)}\end{tabular} & \begin{tabular}[c]{@{}c@{}}\textbf{Alphabit}\\ \textbf{(17)}\end{tabular} & \begin{tabular}[c]{@{}c@{}}\textbf{Rabbit}\\ \textbf{(40)}\end{tabular} & \begin{tabular}[c]{@{}c@{}}\textbf{PractRand}\\ \textbf{(920)}\end{tabular} \\ \midrule
  LFSR NIST 2E 1 & 0                                                   & 0  (2)                                                   & 0                                                 & 0                                                         & 0                                                       & 0                                                     & 0    (0)                                                   \\ \midrule
  LFSR NIST 2E 2 & 0                                                   & 0  (3)                                                   & 0                                                 & 0                                                         & 1                                                       & 0                                                     & 0    (0)                                                   \\ \midrule
  LFSR NIST 2E 3 & 0                                                   & 0  (1)                                                   & 0                                                 & 0                                                         & 0                                                       & 0                                                     & 0   ( 1)                                                   \\ \midrule
  LFSR NIST 2E 4 & 0                                                  & 0  (0)                                                   & 0                                                 & 0                                                         & 1                                                       & 1                                                     & 0    (4)                                                   \\ \midrule
  LFSR NIST 2E 5 & 0                                                  & 0  (0)                                                   & 0                                                 & 0                                                         & 0                                                       & 0                                                     & 0    (3)                                                   \\ \midrule
  \textbf{Total} & \textbf{0}                                         & \textbf{0  (6)}                                          & \textbf{0}                                        & \textbf{0}                                                & \textbf{2}                                              & \textbf{1}                                            & \textbf{0   (8)}                                          \\ \bottomrule
  \end{tabularx}
\end{table}
\unskip

\begin{table}[H]
  \tablesize{\footnotesize}
  \caption{{Statistical} 
 test results for RDSEED as the weak input source to the strong two-source randomness \Circulant extractor. 
  The seed is randomness generated from the NIST Randomness Beacon.
  In cells with multiple entries, failed tests are on the left and suspicious tests (when applicable) are on the right in parentheses.}
  \newcolumntype{C}{>{\centering\arraybackslash}X}
		\begin{tabularx}{\textwidth}{Cccccccc}
  \toprule
  \textbf{RNG}              & \begin{tabular}[c]{@{}c@{}}\textbf{NIST}\\ \textbf{(15)}\end{tabular} & \begin{tabular}[c]{@{}c@{}}\textbf{Diehard}\\ \textbf{(18)}\end{tabular} & \begin{tabular}[c]{@{}c@{}}\textbf{ENT}\\ \textbf{(6)}\end{tabular} & \begin{tabular}[c]{@{}c@{}}\textbf{SmallCrush}\\ \textbf{(15)}\end{tabular} & \begin{tabular}[c]{@{}c@{}}\textbf{Alphabit}\\ \textbf{(17)}\end{tabular} & \begin{tabular}[c]{@{}c@{}}\textbf{Rabbit}\\ \textbf{(40)}\end{tabular} & \begin{tabular}[c]{@{}c@{}}\textbf{PractRand}\\ \textbf{(920)}\end{tabular} \\ \midrule
  RDSEED NIST 2E 1 & 0                                                    & 0  (0)                                                   & 0                                                 & 0                                                         & 0                                                       & 1                                                     & 0    (0)                                                   \\ \midrule
  RDSEED NIST 2E 2 & 0                                                    & 0  (1)                                                   & 0                                                 & 0                                                         & 0                                                       & 1                                                     & 0    (0)                                                   \\ \midrule
  RDSEED NIST 2E 3 & 0                                                    & 0  (2)                                                   & 0                                                 & 0                                                         & 0                                                       & 1                                                     & 0    (1)                                                   \\ 
\bottomrule
  \end{tabularx}
\end{table}

\begin{table}[H]\ContinuedFloat
  \tablesize{\footnotesize}
  \caption{\emph{Cont.}}
  \newcolumntype{C}{>{\centering\arraybackslash}X}
		\begin{tabularx}{\textwidth}{Cccccccc}
  \toprule
  \textbf{RNG}              & \begin{tabular}[c]{@{}c@{}}\textbf{NIST}\\ \textbf{(15)}\end{tabular} & \begin{tabular}[c]{@{}c@{}}\textbf{Diehard}\\ \textbf{(18)}\end{tabular} & \begin{tabular}[c]{@{}c@{}}\textbf{ENT}\\ \textbf{(6)}\end{tabular} & \begin{tabular}[c]{@{}c@{}}\textbf{SmallCrush}\\ \textbf{(15)}\end{tabular} & \begin{tabular}[c]{@{}c@{}}\textbf{Alphabit}\\ \textbf{(17)}\end{tabular} & \begin{tabular}[c]{@{}c@{}}\textbf{Rabbit}\\ \textbf{(40)}\end{tabular} & \begin{tabular}[c]{@{}c@{}}\textbf{PractRand}\\ \textbf{(920)}\end{tabular} \\ \midrule
  RDSEED NIST 2E 4 & 0                                                    & 0  (1)                                                   & 0                                                 & 0                                                         & 0                                                       & 0                                                     & 0    (1)                                                   \\ \midrule
  RDSEED NIST 2E 5 & 0                                                     & 0  (0)                                                   & 0                                                 & 0                                                         & 0                                                       & 0                                                     & 0    (3)                                                   \\ \midrule
  \textbf{Total}   & \textbf{0}                                           & \textbf{0  (4)}                                          & \textbf{0}                                        & \textbf{0}                                                & \textbf{0}                                              & \textbf{3}                                            & \textbf{0    (5)}                                          \\ 
\bottomrule
  \end{tabularx}
\end{table}
\unskip

\begin{table}[H]
  \tablesize{\footnotesize}
  \caption{{Statistical} 
 test results for IDQ Quantis as the weak input source to the strong two-source randomness \Circulant extractor. 
  The seed is randomness generated from the NIST Randomness Beacon.
  In cells with multiple entries, failed tests are on the left and suspicious tests (when applicable) are on the right in parentheses.}
  \newcolumntype{C}{>{\centering\arraybackslash}X}
		\begin{tabularx}{\textwidth}{Cccccccc}
  \toprule
  \textbf{RNG}              & \begin{tabular}[c]{@{}c@{}}\textbf{NIST}\\ \textbf{(15)}\end{tabular} & \begin{tabular}[c]{@{}c@{}}\textbf{Diehard}\\ \textbf{(18)}\end{tabular} & \begin{tabular}[c]{@{}c@{}}\textbf{ENT}\\ \textbf{(6)}\end{tabular} & \begin{tabular}[c]{@{}c@{}}\textbf{SmallCrush}\\ \textbf{(15)}\end{tabular} & \begin{tabular}[c]{@{}c@{}}\textbf{Alphabit}\\ \textbf{(17)}\end{tabular} & \begin{tabular}[c]{@{}c@{}}\textbf{Rabbit}\\ \textbf{(40)}\end{tabular} & \begin{tabular}[c]{@{}c@{}}\textbf{PractRand}\\ \textbf{(920)}\end{tabular} \\ \midrule
  IDQ Quantis NIST 2E 1 & 0                                                    & 0  (0)                                                   & 0                                                 & 0                                                         & 0                                                       & 0                                                     & 0    (0)                                                   \\ \midrule
  IDQ Quantis NIST 2E 2 & 0                                                    & 0  (1)                                                   & 0                                                 & 0                                                         & 0                                                       & 0                                                     & 0    (2)                                                   \\ \midrule
  IDQ Quantis NIST 2E 3 & 0                                                    & 0  (1)                                                   & 0                                                 & 0                                                         & 0                                                       & 0                                                     & 0    (0)                                                   \\ \midrule
  IDQ Quantis NIST 2E 4 & 0                                                    & 0  (0)                                                   & 0                                                 & 0                                                         & 0                                                       & 1                                                     & 0    (2)                                                   \\ \midrule
  IDQ Quantis NIST 2E 5 & 0                                                    & 0  (1)                                                   & 0                                                 & 0                                                         & 0                                                       & 0                                                     & 0    (1)                                                   \\ \midrule
  \textbf{Total}        & \textbf{0}                                           & \textbf{0  (3)}                                          & \textbf{0}                                        & \textbf{0}                                                & \textbf{0}                                              & \textbf{1}                                            & \textbf{0 (5)}                                           \\ \bottomrule
  \end{tabularx}
\end{table}

\section{Physical Randomness Extraction in~Detail}
\label{app:sdi}
\unskip

\subsection{Protocol For Physical Randomness~Extraction}
\label{app:sdira-protocol}
For this physical randomness extraction, we roughly follow the protocol developed in~\cite{foreman2020practical}, with~some adaptations to improve the randomness generation speed. ({{Other} protocols could be used and adapted---for~example~\cite{kessler2020device}---but~additional analysis would need to be performed. Some other protocols allow for fewer assumptions than ours in~\cite{foreman2020practical},\linebreak e.g.,~\cite{chung2014physical,ramanathan2023finite}, but~these require significant classical computation so cannot be implemented on the current hardware}).
This produces a semi-device-independent protocol for randomness amplification using a remote quantum computer, based on Bell tests. 
Roughly speaking, a~Bell test requires a device to be challenged with inputs and then, based on the observed input--output statistics, a~certain amount of entropy can be certified in the outputs.
For a good description of Bell tests, see ``Non-local games'' in~\cite{brunner2014bell}\@. 

The adapted protocol that we use is constructed as follows.
\setlist{nolistsep}
\begin{enumerate}[noitemsep]\label{sdi-prot}
  \item During each of the $n$ rounds, prepare a circuit that generates the GHZ state\linebreak $\frac{1}{\sqrt{2}}(\ket{000} + i\ket{111})$ and measure each qubit with a local $X$ or $Y$ measurement decided by the inputs at that round, which are selected from the set of measurements\linebreak $\{(X,X,X), (X,Y,Y), (Y,X,Y), (Y,Y,X)\}$. 
  Labelling local $X$ and $Y$ measurements as $0$ and $1$, respectively, allows us to write each measurement setting in the set as $(x_i,y_i, x_i \oplus y_i)$, where subscript $i$ denotes the $i$-th round and $x_i, y_i$ are input bits selected using the initial RNG. See {Section~6.3} 
 \textit{{Implementations of Mermin inequality violations on quantum computers}} in~\cite{foreman2020practical}\@.
  \item Run the circuit of round $i \in 1,2,...,n$, recording the measurement settings $x_i,y_i, x_i \oplus y_i$ and measurement outcomes $a_i,b_i,c_i$ of this round.
  \item After $n$ rounds, calculate the observed probability distribution $\Pr(a,b,c | x, y)$. 
  \item Evaluate the Mermin inequality~\cite{mermin1990extreme} value $\mathsf{M_{obs}}$, where
\begin{align} \mathsf{M_{obs}} = E_{0,0,0} - E_{0,1,1} - E_{1,0,1} - E_{1,1,0}   \end{align}
  from the observed probability distribution, where $E_{x,y,x \oplus y}$ denotes the correlator for measurements $(x,y, x \oplus y)$, defined by
\begin{align} E_{x,y,x \oplus y} = \sum_{a\oplus b\oplus c =0} \Pr(a,b,c \:|\: x, y, x \oplus y ) - \sum_{a\oplus b\oplus c =1} \Pr(a,b,c \:|\: x, y, x \oplus y ) \end{align}
  \item Reduce $\mathsf{M_{obs}}$ to account for finite statistics using the H\"{o}effding inequality, using the relationship between $\mathsf{M_{obs}}$ and the `losing probability', found at the beginning of Appendix A.2 of~\cite{foreman2020practical}\@. Let $\epsilon_{\rm est}$ be the estimation error and $\mathsf{M}$ be the true (asymptotic) value of the Mermin inequality for some I.I.D. quantum device; then, we find $\mathsf{M_{adj}}$ such that $\Pr(\mathsf{M_{adj}} > \mathsf{M}) \leq \epsilon_{\rm est}$ by defining
\begin{align}
      \mathsf{M_{adj}} = \mathsf{M_{obs}} - 16 t.
  \end{align} and
\begin{align}
      \epsilon_{\rm est} = \exp(-2t/n)
  \end{align} for $t > 0$.
  \item Based on the adjusted value $\mathsf{M_{adj}}$, evaluate the amount of min-entropy in the measurement outcomes of the quantum device. This is performed using the analytic expression in~\cite{woodhead2018randomness}, which applies to two output bits per round. For~details, see {Section~4.3} \textit{{Quantum devices, Bell tests, and~guessing probabilities}} \cite{foreman2020practical}. Note that the~relationship between the guessing probability and min-entropy can be found in Appendix A.3, Equation~(28) of~\cite{foreman2020practical}\@.
  \item Take two of the three output {bits} 
 (e.g.,\ $\mathbf{a},\mathbf{b}$, discard $\mathbf{c}$) for randomness extraction.
  \item Perform strong two-source randomness extraction using the quantum computer outputs $(\mathbf{a},\mathbf{b})$ with a fresh string of randomness from an RNG, if~the sum of the min-entropies of each bit string is high enough for extraction. The~output is a near-perfect bit string, which we call the seed.
  \item Repeatedly perform strong seeded extraction using the generated seed and fresh strings from the initial RNG, concatenating the output following the same logic as level 3, i.e., two-source extraction.
\end{enumerate}

For the extractor implementation, we again use the $\Circulant$ extractor, and~steps (8) and (9) can be viewed as analogous to level 3, where the source $Y$ is instead generated by the described quantum process. We used the $\mathsf{H}$1-1 Quantinuum ion-trap quantum computer as our quantum device, executing $1.8 \times 10^6$ circuits to obtain a weakly random seed of $3.6 \times 10 ^ 6$ bits. 
This experiment took approximately 33.5 h of quantum computing time. 
We obtain $\mathsf{M_{obs}} = 3.83 \rightarrow \mathsf{M_{adj}} \approx 3.75 \rightarrow \alpha_{\mathsf{Q}} \geq 0.518$, where $\alpha_{\mathsf{Q}}$ is the min-entropy rate of the quantum computer outputs $(\mathbf{a},\mathbf{b})$. 
We note that the min-entropy of the LSFR was too low to perform physical extraction, as~a necessary condition is $\alpha_{\mathsf{RNG}} > 1 - \alpha_{\mathsf{Q}}$.

\subsection{Full~Results}
\vspace{-6pt}
\begin{table}[H]
  \caption{Statistical test results for RDSEED as the weak input source to the physical randomness extractor hierarchy level, implemented with the \Circulant extractor. 
  The seed is randomness generated using the semi-device-independent randomness amplification protocol outlined in \Cref{app:sdira-protocol}\@.
  In cells with multiple entries, failed tests are on the left and suspicious tests (when applicable) are on the right in parentheses.}
  \newcolumntype{C}{>{\centering\arraybackslash}X}
		\begin{tabularx}{\textwidth}{Cccccccc}
  \toprule
  \textbf{RNG}              & \begin{tabular}[c]{@{}c@{}}\textbf{NIST}\\ \textbf{(15)}\end{tabular} & \begin{tabular}[c]{@{}c@{}}\textbf{Diehard}\\ \textbf{(18)}\end{tabular} & \begin{tabular}[c]{@{}c@{}}\textbf{ENT}\\ \textbf{(6)}\end{tabular} & \begin{tabular}[c]{@{}c@{}}\textbf{SmallCrush}\\ \textbf{(15)}\end{tabular} & \begin{tabular}[c]{@{}c@{}}\textbf{Alphabit}\\ \textbf{(17)}\end{tabular} & \begin{tabular}[c]{@{}c@{}}\textbf{Rabbit}\\ \textbf{(40)}\end{tabular} & \begin{tabular}[c]{@{}c@{}}\textbf{PractRand}\\ \textbf{(920)}\end{tabular} \\ \midrule
  RDSEED PE 1 & 0                                                   & 0  (1)                                                   & 0                                                 & 0                                                         & 0                                                       & 0                                                     & 0   (0)                                                   \\ \midrule
  RDSEED PE 2 & 0                                                   & 0  (0)                                                   & 0                                                 & 0                                                         & 0                                                       & 0                                                     & 0   (2)                                                   \\ \midrule
  RDSEED PE 3 & 0                                                   & 0  (0)                                                   & 0                                                 & 0                                                         & 0                                                       & 0                                                     & 0   (0)                                                   \\ 
\bottomrule
  \end{tabularx}
\end{table}

\begin{table}[H]\ContinuedFloat
  \caption{\emph{Cont.}}
  \newcolumntype{C}{>{\centering\arraybackslash}X}
		\begin{tabularx}{\textwidth}{Cccccccc}
  \toprule
  \textbf{RNG}              & \begin{tabular}[c]{@{}c@{}}\textbf{NIST}\\ \textbf{(15)}\end{tabular} & \begin{tabular}[c]{@{}c@{}}\textbf{Diehard}\\ \textbf{(18)}\end{tabular} & \begin{tabular}[c]{@{}c@{}}\textbf{ENT}\\ \textbf{(6)}\end{tabular} & \begin{tabular}[c]{@{}c@{}}\textbf{SmallCrush}\\ \textbf{(15)}\end{tabular} & \begin{tabular}[c]{@{}c@{}}\textbf{Alphabit}\\ \textbf{(17)}\end{tabular} & \begin{tabular}[c]{@{}c@{}}\textbf{Rabbit}\\ \textbf{(40)}\end{tabular} & \begin{tabular}[c]{@{}c@{}}\textbf{PractRand}\\ \textbf{(920)}\end{tabular} \\ \midrule
  RDSEED PE 4 & 0                                                  & 0  (1)                                                   & 0                                                 & 0                                                         & 0                                                       & 0                                                     & 0   (1)                                                   \\ \midrule
  RDSEED PE 5 & 0                                                 & 0  (0)                                                   & 0                                                 & 0                                                         & 0                                                       & 1                                                     & 0   (0)                                                   \\ \midrule
  \textbf{Total}  & \textbf{0}                                         & \textbf{0  (2)}                                          & \textbf{0}                                        & \textbf{0}                                                & \textbf{0}                                              & \textbf{1}                                            & \textbf{0  (3)}                                          \\ 
\bottomrule
  \end{tabularx}
\end{table}
\unskip

\begin{table}[H]
  \caption{Statistical test results for IDQ Quantis as the weak input source to the physical randomness extractor hierarchy level, implemented with the \Circulant extractor. 
  The seed is randomness generated using the semi-device-independent randomness amplification protocol outlined in \Cref{app:sdira-protocol}\@.
  In cells with multiple entries, failed tests are on the left and suspicious tests (when applicable) are on the right in parentheses.}
\newcolumntype{C}{>{\centering\arraybackslash}X}
		\begin{tabularx}{\textwidth}{Cccccccc}
  \toprule
  \textbf{RNG}              & \begin{tabular}[c]{@{}c@{}}\textbf{NIST}\\ \textbf{(15)}\end{tabular} & \begin{tabular}[c]{@{}c@{}}\textbf{Diehard}\\ \textbf{(18)}\end{tabular} & \begin{tabular}[c]{@{}c@{}}\textbf{ENT}\\ \textbf{(6)}\end{tabular} & \begin{tabular}[c]{@{}c@{}}\textbf{SmallCrush}\\ \textbf{(15)}\end{tabular} & \begin{tabular}[c]{@{}c@{}}\textbf{Alphabit}\\ \textbf{(17)}\end{tabular} & \begin{tabular}[c]{@{}c@{}}\textbf{Rabbit}\\ \textbf{(40)}\end{tabular} & \begin{tabular}[c]{@{}c@{}}\textbf{PractRand}\\ \textbf{(920)}\end{tabular} \\ \midrule
  IDQ Quantis PE 1 & 0                                                   & 0  (1)                                                   & 0                                                 & 0                                                         & 0                                                       & 0                                                     & 0  (1)                                                   \\ \midrule
  IDQ Quantis PE 2 & 0                                                   & 0  (0)                                                   & 0                                                 & 0                                                         & 0                                                       & 1                                                     & 0  (2)                                                   \\ \midrule
  IDQ Quantis PE 3 & 0                                                   & 0  (0)                                                   & 1                                                 & 0                                                         & 0                                                       & 0                                                     & 0  (3)                                                   \\ \midrule
  IDQ Quantis PE 4 & 0                                                   & 0  (2)                                                   & 0                                                 & 0                                                         & 0                                                       & 0                                                     & 0  (0)                                                   \\ \midrule
  IDQ Quantis PE 5 & 0                                                   & 0  (0)                                                   & 0                                                 & 0                                                         & 0                                                       & 1                                                     & 0  (1)                                                   \\ \midrule
  \textbf{Total}       & \textbf{0}                                          & \textbf{0  (3)}                                          & \textbf{1}                                        & \textbf{0}                                                & \textbf{0}                                              & \textbf{2}                                            & \textbf{0  (7)}                                          \\ \bottomrule
  \end{tabularx}
\end{table} 

\begin{adjustwidth}{-\extralength}{0cm}

\reftitle{References}



\PublishersNote{}
\end{adjustwidth}
\end{document}